# Topotaxial Mutual-Exchange Growth of Magnetic Zintl Eu$_3$In$_2$As$_4$ Nanowires with Axion Insulator Classification

Man Suk Song[1], Lothar Houben[2], Yufei Zhao[1], Hyeonhu Bae[1], Nadav Rothem[3], Ambikesh Gupta[1], Binghai Yan[1], Beena Kalisky[3], Magdalena Zaluska-Kotur[4], Perla Kacman[4], Hadas Shtrikman[1]*, Haim Beidenkopf[1]*

[1]Department of Condensed Matter Physics, Weizmann Institute of Science; Rehovot, 7610001, Israel.
[2]Department of Chemical Research Support, Weizmann Institute of Science; Rehovot, 7610001, Israel.
[3]Department of Physics and Institute of Nanotechnology and Advanced Materials, Bar-Ilan University; Ramat Gan, 5290002, Israel.
[4]Institute of Physics, Polish Academy of Sciences; Warsaw, PL-02-668, Poland.

*Corresponding author. Email: hadas.shtrikman@weizmann.ac.il (H.S.); haim.beidenkopf@weizmann.ac.il (H.B.)

**Abstract:** Nanomaterials bring to expression unique electronic properties that promote advanced functionality and technologies. Albeit, nanoscale growth presents paramount challenges for synthesis limiting the diversity in structures and compositions. Here, we demonstrate solid-state topotactic exchange that converts Wurtzite InAs nanowires into Zintl phase Eu$_3$In$_2$As$_4$ nanowires. In situ evaporation of Eu and As over InAs nanowire cores in molecular beam epitaxy results in mutual exchange of Eu from the shell and In from the core. A continuous Eu$_3$In$_2$As$_4$ shell thereby grows that gradually consumes the InAs core and converts it into a single phase Eu$_3$In$_2$As$_4$ nanowire. Topotaxy, which facilitates the mutual exchange, is supported by the substructure of the As matrix which is similar across the Wurtzite InAs and Zintl Eu$_3$In$_2$As$_4$. We provide initial evidence of an antiferromagnetic transition at $T_N \sim 6.5$ K in the Zintl phase Eu$_3$In$_2$As$_4$ nanowires. Ab initio calculation confirms the antiferromagnetic state and classifies Eu$_3$In$_2$As$_4$ as a $C_2T$ axion insulator hosting both chiral hinge modes and unpinned Dirac surface states. The topotactic mutual-exchange growth of Zintl Eu$_3$In$_2$As$_4$ nanowires thus enables the exploration of intricate magneto-topological states of nanomaterials. Moreover, it may open the path for topotactic mutual-exchange synthesis of nanowires made of other exotic compounds.

Material science has been profoundly influenced by the remarkable contributions of diverse material growth techniques. These techniques have revolutionized the way we synthesize and engineer materials with great variety and tailored properties, enabling groundbreaking advancements in scientific and technological domains. Advanced growth methodologies enable control not only over composition and structure but also over size, geometry and dimensionality. Beyond three-dimensional crystals, these include also quasi-two-dimensional thin films, interfaces and heterostructures, quasi-one-dimensional structures such as nanowires (NWs) and nanotubes, and zero-dimensional quantum dots. We focus here on NWs that have opened new horizons for manipulating matter at the nanoscale, enabling innovative applications across a broad range of disciplines[1–4]. NWs are synthesized using various growth methods, each offering unique advantages and control over their properties[5–7]. These include vapor-phase methods from gaseous precursors[8], solution-based methods from liquid precursors[9], Electrochemical methods[10], Template-assisted growth from a porous template[11,12] and seed-mediated growth[13]. Many of those have found realization also in ultra-pure molecular beam epitaxial (MBE) growth[5–7,13].

Still, the strict geometrical constraints imposed by the extreme aspect ratio of the NWs limit the variety of compounds that can be readily grown in this form, hence limiting their functionality. Here, we adopt the approach of topotactic conversion[14–18] to solid-state growth of NWs of exotic composition and properties. Topotaxy refers to the controlled alteration of a crystal's composition while either fully or partially preserving a substructure of the original crystal in structure and orientation. This process allows for the transformation of one material into another with minimal disruption to the arrangement of atoms or ions. Topotactic conversion has significant implications for various applications, from catalysis and energy storage to electronics and optics, as it enables the creation of advanced materials with precise modifications while retaining their structural integrity. Topotaxy is particularly valuable in the realm of nanomaterials, where NWs and other nanostructures can be transformed into new materials with tailored characteristics[19–21]. Topotactic conversion of NWs has been demonstrated in limited material families using several techniques. Prime examples include the conversion of NWs from metal oxide to metal[22], conversion of various chalcogenide NWs[17,18] and reversible conversion during charging and discharging of perovskite NWs based batteries[23,24].

The overwhelming majority of previous demonstrations of topotactic conversion was achieved in solution-based environments that act as an abundant ionic reservoir for the exchange process, AB + C → AC + B, that takes place within the NW. Here, we report an in-situ vapor-solid topotactic conversion of NWs grown by MBE. It is the first demonstration of a conversion of III-V InAs NWs in general, and particularly into Zintl $Eu_3In_2As_4$. In the absence of ionic reservoir we reveal a unique mutual exchange process, AB + BC → ABC (A=Eu, B=As, C=In), whereby the $Eu_3In_2As_4$ phase simultaneously progresses inwards into the In-rich InAs core and outwards over the Eu-rich shell. To obtain the single phase $Eu_3In_2As_4$ crystal, Eu and In are mutually exchanged. We find that topotaxy is essential to this process and established through the As matrix which is similar across the Wurtzite (WZ) InAs and Zintl $Eu_3In_2As_4$ structures.

We first describe the MBE growth protocol and underlying mutual-exchange mechanism that enabled us to grow $Eu_3In_2As_4$ NWs and identify their crystallographic structure. At the initial MBE growth stage we use the well-established gold (Au) assisted vapor-liquid-solid (VLS) method to grow WZ InAs NWs (see methods for details) both on a (001) InAs and a (111)B terminated substrate[13,25]. Since the NWs grow along the <111> direction, the former results in reclining NWs and the latter in NWs normal to the substrate surface. The InAs cores are about 50 nm in diameter and 4-8 microns long. Then, the In source is closed and cooled down, and Eu and As are being evaporated for 1-4 hours. A scanning electron microscope (SEM) image of the resulting NWs, grown on a (001) substrate, is shown in Fig. 1a. The NWs seem to have a certain degree of roughness. It is better resolved in the zoomed in SEM image in Fig. 1b showing polycrystalline-like structure along the NW axis. We show below that this roughness originates from alternation between two distinct crystallographic orientations.

We examine the NW's stoichiometry and internal structure by atomically resolved transmission electron microscopy (TEM). A representative image is given in Fig. 1c, showing crystallites of a perfectly ordered lattice over the WZ InAs core. The crystallites grow both outwards as a shell as well as inwards penetrating into the NW core. We further use energy dispersive X-ray spectroscopy (EDS) to resolve the elemental composition. Remarkably, even though the core grows in the complete absence of Eu and the shell grows in the complete absence of In, in the crystallites that form we find a uniform distribution of In and Eu (Fig. 1d, see separate Eu and In imaging in Supplementary Fig. 1) alongside As (Fig. 1e). Such mutual exchange of In from the core and Eu from the shell is common to all the NWs that we have grown under such protocol. As both the TEM and EDS images show, the penetration of Eu ions into the core and In atoms into the shell does not form a gradient of concentrations typical of diffusion processes. In contrast, an atomically sharp phase boundary forms between InAs on the core side and Eu-In-As on the shell side that progressively propagates inwards via the mutual exchange process of Eu and In across it.

From the analysis of the relative EDS intensities of the three elements, which are shown in Supplementary Fig. 1e, we determine the stoichiometry of the shells as $Eu_3In_2As_4$. We found this exact stoichiometry at all ternary shells of all NWs that we have examined (Supplementary Fig. 2). High resolution scanning TEM, shown in Fig. 1f, displays the perfectly ordered structure of the $Eu_3In_2As_4$ crystallites. TEM-EDS, shown in Fig. 1g and h, enables us to identify the material as a Zintl phase with the orthorhombic space group $Pnnm$[26]. Its unit cell is shown in Fig. 1i as well as overlaid on the scanning TEM image in Fig. 1f (see also Supplementary Fig. 3). $Eu_3In_2As_4$ consists of quasi-one dimensional chains of $In_2As_4$ embedded in a matrix formed by herringbone-like Eu triplets (illustrated in Supplementary Fig.11). During the uptake of Eu, the corner-shared InAs tetrahedral coordination in WZ InAs breaks into pairs of slightly distorted InAs tetrahedra that share one of their tetrahedral edges. These $In_2As_4$ pairs connect to continuous $[In_2As_4]_n$-chains by sharing tetrahedral corners. The chains are poly-anionic with their $[In_2As_4]^{6-}$ pair units, which are aligned with the $c$-axis of the orthorhombic $Eu_3In_2As_4$ phase. Triplets of $Eu^{2+}$ cations separate the poly-anionic $[In_2As_4]_n$ chains. This type of anionic chain formation is also characteristic of the isostructural Zintl phases of $Sr_3In_2P_4$ and $Ca_3In_2As_4$[27].

Yet, we find that the mutual exchange process is not a virtue only of the exchanging elements, In from the core and Eu from the shell, but depends crucially on the structure through which they diffuse as well. We find that the WZ structure has a remarkable property – its As sublattice is almost identical to the As sublattice in $Eu_3In_2As_4$, as demonstrated in Fig. 2a. Hence, as the $Eu_3In_2As_4$ grains grow, Eu and In mutually exchange, retaining a similar As skeleton that exists in the InAs core and continues to grow in the shell. Such similarity between part of the parent and resulting crystals is termed topotaxy rendering the growth process we have identified as topotaxial mutual-exchange growth. Topotaxy renders the exchange process selective to the initial growth structures. In our previous attempts[28] with depositing a Eu, As, and In shell over either WZ or zincblende (ZB) NW cores we have only achieved a mosaic structure of EuAs sheets embedded within polycrystalline InAs. Here we show the growth of a crystalline Zintl $Eu_3In_2As_4$ phase from WZ InAs by deposition Eu and As but no In. We have found that following a similar protocol with ZB NW cores results in the exchange of a distinct Zintl $Eu_5In_2As_6$ shell which is topotactic with the As substructure of ZB (to be reported elsewhere), demonstrated in Supplementary Fig.4. This highlights how topotaxy selectively dictates the end products by their initial As substructure as well as the potential richness of the method.

With this, we also resolve the origin for the roughness of the $Eu_3In_2As_4$ shell. The conversion occurs throughout the facets of the original InAs core. With atomic resolution in the scanning TEM, we find a single resulting $Eu_3In_2As_4$ phase in composition and structure. However, we observe precisely two distinct orientational relationships between the orthorhombic $Eu_3In_2As_4$ and the underlying InAs, labeled as type I and type II in Fig. 2. In type I (Figs. 2b and c), the orientation between the InAs and the $Eu_3In_2As_4$ is described by $\langle 10\bar{1}0 \rangle_{InAs} \parallel \langle 001 \rangle_{EuInAs}$ and the interfacial planes are $(01\bar{1}0)_{InAs} \parallel (\bar{2}30)_{EuInAs}$. The atomistic model in Fig. 2c shows the $[In_2As_4]_n$ chains along the $c$-axis of the $Eu_3In_2As_4$ Zintl phase aligned with the viewing direction. The connecting $Eu^{2+}$ ions between the chains are well separated as homonuclear columns. The key feature of the conversion (see illustration in Supplementary Fig.11) is a transformation of the polyhedral network of tetrahedral coordination sites and octahedral sites between the InAs and $Eu_3In_2As_4$. The As sub-lattice in WZ InAs forms a hexagonal lattice with four tetrahedral sites and two octahedral sites per unit cell, labeled in Fig. 2c as $T+$, $T-$ and $O$, respectively. In InAs, only half of the tetrahedral $T+$ sites are occupied by In, while the other $T-$ half and the octahedral $O$ sites are vacant. One of the main characteristics of a topotaxial transformation is the diffusion of components – namely the co-diffusion of In and Eu in our case. The mass transport is presumably by a vacancy diffusion mechanism of In and Eu. The In ions occupy vacant $T-$ sites in parallel to the co-diffusion, whereby the InAs tetrahedral pairs and the characteristic InAs chains are formed.

For the type I interface both the As sub-lattice, and with it the polyhedral network of tetrahedral and octahedral sites, undergo minor distortions that are merely local. There is a one-to-one relationship between an As lattice site in the InAs and its counterpart in the $Eu_3In_2As_4$. For example, in Fig. 2c the As sites 1-7 marked in InAs are corner sites of coordination tetrahedra and octahedra, while their counterparts 1'-7' in $Eu_3In_2As_4$ are the corner sites of coordination polyhedra of the identical type. A minor increase of the lattice volume per As atom during the transformation from 57 $Å^3$ in InAs to 62.2 $Å^3$ in $Eu_3In_2As_4$ simply reflects the larger ionic radius of Eu compared with In. Occasionally we observe stacking faults with varying concentrations that locally modify the typical herringbone-like crystal structure. They occur during the growth of type I interface on the (010) planes of $Eu_3In_2As_4$, by mirror flip of the orientation of the InAs tetrahedra to opposite faces of the EuAs octahedra along the (100) direction. To our observation, the stacking fault density is potentially related to the roughness of the phase boundary between the $Eu_3In_2As_4$ and InAs. The stacking faults are most evident in the In and Eu sublattices. The As sublattice shows a minor distortion in the form of a small rigid body shift across the stacking fault, the hexagonal As lattice remains continuous (Supplementary Fig. 8).

For type II in Figs. 2d and e, the orientation relationship between InAs and $Eu_3In_2As_4$ is $\langle 1\bar{2}10 \rangle_{InAs} \parallel \langle 100 \rangle_{EuInAs}$. In this case, the interface between InAs and $Eu_3In_2As_4$ is inclined by approximately 109° with respect to the viewing direction. This inclination produces the seemingly gradual transition between InAs and $Eu_3In_2As_4$ in this TEM viewing direction. Unlike the interface of type I, the InAs chains in type II crystals are not aligned with a $\langle 10\bar{1}0\rangle_{\text{InAs}}$ direction in InAs. Instead, they are running diagonally in the $(1\bar{2}10)$ plane as seen in the TEM image in Fig. 2d and the corresponding modeling in Fig. 2e. Since the parent phase and the resulting phase of the topotaxial transformation are the same as for type I, we are able to assign a similar direct relationship between the As sites forming corners of the polyhedral lattices. However, there is an additional octahedrally coordinated layer of EuAs at the interface between the InAs and the $Eu_3In_2As_4$ that causes a symmetry flip. The circle marker in Fig. 2e marks one of these additional octahedra that is face-sharing with a neighboring tetrahedral site in the InAs. We note that this is precisely a symmetry element in cubic InAs, while in hexagonal InAs all octahedral sites are edge-sharing with a tetrahedral site. Even though we do not have an edge-on view of the interface between the InAs and the $Eu_3In_2As_4$ we can, for this case, construct a possible interface type. The continuation of a face-sharing tetrahedra-octahedra plane leads to a possible solution where the interfacial planes are given by $(20\bar{2}1)_{\text{InAs}} \parallel (\bar{1}40)_{\text{EuInAs}}$.

From the preservation of the polyhedral network defined by the As lattice type I and II we draw a commonality: the polyhedral corners of InAs tetrahedra and octahedra emerge as the connecting edges of InAs tetrahedra in the $[In_2As_4]^{6-}$ pairs and separating octahedral coordination sites of $Eu^{2+}$ that separate the chains. There are two of these chains running parallel with alternating rotation around the $c$-axis of the $Eu_3In_2As_4$. For the type I interface these chains are running orthogonal to the growth axis of the initial InAs nanowire. For the type II interface they are inclined by approximately 29° with respect to the growth axis.

Based on high-resolution EDS-TEM imaging we model full grains of both types and the interfaces they form with the InAs core. The type I grain from Fig. 1c is shown in Fig. 2f. The As sub-structure of that grain and core is displayed in the inset demonstrating the continuous rows that cross the interface. Figure 2g shows a full atomic model of the type II grain and core from the inset image. The two orientations are present in the initial stages of the topotaxial transformation, where we find that type II crystallites are about twice more common than type I (see statistics in Supplementary Fig. 9). They prevail when the conversion occurs throughout the entire InAs NW, eventually leading to a multidomain structure in fully converted NWs. So far we have not found a protocol that will allow one type to dominate and yield a single crystalline $Eu_3In_2As_4$ structure. However, slight energetic differences and strain render it potentially feasible, calling for further investigation along this line.

Next, we examine the parameter phase space that supports topotaxial exchange growth of the $Eu_3In_2As_4$ NWs. We have examined both (001) and (111) aligned InAs substrates, resulting in NW cores that are either reclining or normal to the substrate, respectively. On the reclining NWs we find growth of $Eu_3In_2As_4$ half shell on the surfaces that face the evaporation cells. This enables easy visualization of the interface between the progressing $Eu_3In_2As_4$ shell and the consumed InAs core. However, the reclining NWs were found to be less stable to elevated temperatures. On the vertical NWs we find a full $Eu_3In_2As_4$ shell that progresses radially inwards. This growth is both more efficient and more robust to elevated temperatures but the core-shell interface of it is internal and its detection requires TEM imaging.

We prolonged the shell growth from 1 to 4 hours keeping all other parameters fixed, such as the substrate temperature of 420 °C and the Eu and As fluxes. We examined this with reclining NW cores where the $Eu_3In_2As_4$ progression can be readily imaged. We show in Fig. 3a with scanning TEM-EDS mapping that as the growth time was extended from one to three hours the $Eu_3In_2As_4$ shell thickened while it progressively consumed the InAs core. Again we found that along this process In and Eu mutually exchanged while maintaining a sharp interface between the core and the shell rather than developing a gradient of concentrations typical of diffusion. After four hours of growth period the $Eu_3In_2As_4$ shell consumed the entire 50 nm diameter InAs core leaving no residual In available. As a result, in the absence of In, an Eu-As shell began to form.

We further examined in Fig. 3b the dependence on substrate temperature keeping the growth duration fixed at 2 hours. For this we used vertical NWs grown on (111) substrates that are more stable at elevated temperatures. At the low temperature of 370 °C exchange did not seem to occur. At the end of that growth we found a Eu-As shell over an InAs core without any clear exchange of Eu and In as resolved by scanning TEM-EDS. At a slightly higher temperature of 400 °C overlap in the Eu and In distributions is clearly resolved as the mutual-exchange onsets. The exchange process was robust across a temperature interval of about 100 °C, becoming more efficient with increasing temperature. At 460 °C we identified a different growth pattern in which the NW segregates into a necklace of the type I and type II crystallites [29]. At the same temperature range the base of the NWs narrowed along the exchange growth process to the extent that at 460 °C many of the NWs began to collapse. This is a result of the temperature gradient along the NW axis in which the temperature is higher closer to the substrate (see Supplementary Fig. 7). We note that this is a rare direct observation of the otherwise assumed temperature gradient that exists along the NW as it grows from the temperature monitored substrate. Thus the growth window for topotaxial exchange growth of $Eu_3In_2As_4$ is at 370-460 °C for a duration of 1-4 hours for initial InAs NWs of 50 nm diameter. It is remarkable that the growth phase diagram for topotactic exchange growth of $Eu_3In_2As_4$ overlaps with the growth and stability phase diagram of InAs NWs enabling this novel growth method of Zintl NWs.

We find qualitative agreement between the experimental phase diagram and molecular dynamics simulations. We simulated WZ InAs NWs at various temperatures in the presence of impinging Eu atoms. As shown in Fig. 3c, at low temperatures of 450 K the Eu (green) accumulates on the surface of the InAs NW (magenta-blue). In contrast, at slightly higher temperature of 650 K Eu and In begin to mutually exchange. At yet higher temperatures the Eu atoms disassociate from the InAs surface. Comparing WZ to ZB structures of the InAs NWs in simulations (see Supplementary Fig. 10) confirms that mutual exchange at these conditions is unique to WZ while in ZB NWs we find signatures for the formation of a separate mosaic Eu shell. This further confirms the crucial role of topotaxy in the exchange growth process.

In this final part we provide initial characterization of the coupled magnetic and electronic phase diagrams of the $Eu_3In_2As_4$ NWs. Bulk single crystals of this Zintl compound have been only recently grown in a flux method[26], never before in MBE nor in the form of NWs, and its electronic and magnetic properties have hardly been characterized. Band structure calculations find it is a narrow gap semiconductor and predict it to be antiferromagnetic (AFM) at low temperatures[30]. Similar Eu-containing compounds, such as $Eu_3In_2P_4$ and $EuIn_2As_2$, become AFM below $T_N=14.5$[31] and 16 K[32], respectively. The latter has drawn interest because of its classification as a topologically non-trivial axion insulator and large magneto-topological tunability by means of temperature and magnetic field[32–34].

We thus provide here an initial characterization of the low temperature magnetic ground state of $Eu_3In_2As_4$. We used a scanning superconducting quantum interference device (SQUID) to image the DC magnetization, M, and AC susceptibility, $\chi$, facilitated by an on-chip coil (see methods) of the $Eu_3In_2As_4$ NWs. Corresponding optical and susceptibility images of a few NWs distributed on a nonmagnetic $Si/SiO_2$ substrate are shown in Fig. 4a. We did not detect any DC magnetization signal from the Eu$_3$In$_2$As$_4$ NWs (see Supplementary Fig. 13), but whenever we scanned over a NW we found a strong paramagnetic signal in AC susceptibility. Lack of DC magnetization puts a strict limit on the strength of magnetization associated with a ferromagnetic (FM) order down to our sensitivity [35], but the strong paramagnetic signal signifies the presence of magnetic moments. We examined the dependence of the AC susceptibility across a NW on temperature, shown in Fig. 4b. The paramagnetic susceptibility signal grew as we lowered the temperature. The maximal signal from the NW above the background, $\Delta\chi$, displayed in Fig.4c, shows that below 6.5 K this trend saturates. This commonly indicates a magnetic phase transition from a paramagnetic state to an AFM order. The transition is also evident in the inset that shows the temperature dependence of the inverse susceptibility. Intriguingly, fitting a Curie-Weiss linear temperature dependence yields a positive Weiss constant, $\Theta$, which is typical of FM correlations rather than AFM ones. This peculiarity is shared by several AFM Zintl compounds such as Eu$_3$In$_2$P$_4$, EuMn$_2$P$_2$, and EuSnP[31,36,37] and commonly ascribed to short range FM correlations within the paramagnetic phase. The Neel temperature we found is somewhat lower than in similar Eu-containing Zintl compounds, which may indicate effects of the quasi-one dimensionality or the unusual surface to bulk ratio met in NWs. We further map the magnetic phase diagram by measuring the susceptibility of Eu$_3$In$_2$As$_4$ nanowires dispersed over a Si substrate (see details in Supplementary Fig.14) in a magnetic properties measurement system (MPMS). With increasing magnetic field the cusp in the susceptibility shifts to lower temperatures and flattens. This is captured in Fig.4d which shows its derivative with respect to temperature, d$\chi$/dT(H, T), where the Neel transition is hallmarked by zero slope (see raw susceptibility in Supplementary Fig.14). Curie-Weiss yields consistently a positive Weiss constant, as shown by the inset.

The coexistence of magnetic order and the strong spin-orbit coupling in Eu$_3$In$_2$As$_4$ promotes exotic topological phases. By ab initio calculations we indeed predict the exciting formation of a unique axion insulator phase[38] that for the first time hosts coexisting chiral hinge states and Dirac surface states. We complemented the magnetization measurements with detailed ab initio calculations (see methods) of the magnetic ground states by considering paramagnetic, AFM and FM orders, as shown in Supplementary Fig. 18. Two types of AFM orders were considered (sketched in Supplementary Fig. 16): C-type that indicates AFM order within planes and FM coupling among them along the perpendicular axis, and A-type that denotes the AFM stacking of FM planes[39]. Antiparallel spin alignment between the nearest Eu atoms lowers the electronic energy. Hence, among orders considered C-type AFM was identified as the magnetic ground state of bulk Eu$_3$In$_2$As$_4$, in agreement with our AC susceptibility measurements. Yet, the real AFM order may be more intricate than those considered here, as seen in other Zintl compounds[40]. The magnetic anisotropy energy is negligible for spins along *a*, *b*, and *c* axes. The energy difference between the AFM and FM ground states is only 2 meV. This could explain the FM correlation indicated by the Weiss constant. It also indicates that spins could be polarizable into FM order by a magnetic field. This will enable the exploration of the corresponding magneto-topological phase diagram that we discuss next.

Within its paramagnetic state, bulk Eu$_3$In$_2$As$_4$ is a narrow gap semiconductor with a direct gap of about 50 meV at the $\Gamma$ point (see Supplementary Fig. 19). Other than those bands at $\Gamma$ the band structure is clean of states over almost 0.5 eV above and below the Fermi energy. Because the crystal is inversion symmetric, it is sufficient to examine the parity of occupied bands at time-reversal invariant momenta (TRIM) in order to determine the electronic topological classification ($\mathbb{Z}_4$ index):

$$\mathbb{Z}_4 = \frac{1}{2} \sum_{a=1}^{8} \sum_{n=1}^{n_{occ}} [1 + \xi_n(\Lambda_a)] \mod 4$$

where $n_{occ}$ is the number of occupied bands, $\xi_n(\Lambda_a)$ is the parity eigenvalue (±1) of the nth band at the TRIM $\Lambda_a$. $\mathbb{Z}_4 = 0$ indicates a trivial insulator while $\mathbb{Z}_4 = 1, 3$ denotes a Weyl semimetal. $\mathbb{Z}_4 = 2$ corresponds to an axion insulator with a quantized topological magnetoelectric response ($\theta = \pi$).

For the paramagnetic phase of $Eu_3In_2As_4$ we find $\mathbb{Z}_4 = 0$ classifying it as a topologically trivial semiconductor. In contrast, we found that within the C-AFM phase $Eu_3In_2As_4$ is a nontrivial axion insulator. For the C-AFMb phase for instance, where spins align along the *b* axis as in Fig. 4e, we find a reduced energy gap of about 9 meV (Fig. 4f) with a topological index of $\mathbb{Z}_4 = 2$. The spin-orbit coupling results in band inversion relative to the paramagnetic trivial gap shown by the orbital texture in Fig.4g (the same holds also for C-AFMa and C-AFMc, see SI for details). This classifies $Eu_3In_2As_4$ within its C-AFM order as an axion insulator. Accordingly, we find *bc* and *ac* surfaces are gapped and a one-dimensional chiral hinge mode is found among the gapped surfaces as sketched in Fig. 4h and calculated in Figs. 4i and j, respectively. Unpinned Dirac bands are found on *ab* surfaces, calculated in Fig. 4k. The Dirac nodes there are locally protected by the combined $C_2T$ symmetry, where $C_2$ is a two-fold rotation around the *c* axes and *T* represents time-reversal symmetry[41,42]. Even though ab initio calculations are for a bulk-like slab geometry, by examining the thickness dependence we conclude that the topological bulk gap remains inverted down to a thickness of 28 nm along the *a* direction and 112 nm along the *b* which are both compatible with the dimensions of the $Eu_3In_2As_4$ NWs. Accordingly, the multifaceted NW structure we report here, made of $Eu_3In_2As_4$ crystallites of two orientations, would support electronically a network of chiral modes bound to the hinges and running along the NW axis. Such chiral modes are topologically protected from backscattering. We note that rotating the spin from *b* to *a* or *c* axes will modify the hinge states and Dirac states sensitively as dictated by the $C_2T$ symmetry change.

We have further considered the electronic nature of $Eu_3In_2As_4$ within the FM state, accessible by the application of a magnetic field. We find $\mathbb{Z}_4 = 1$ for moments directed along any of the prime axes. This signifies a Weyl semimetal state with Fermi arc boundary modes on its surfaces that correspond to Weyl nodes in the bulk, sketched in Fig. 4l and calculated in Fig. 4m. Aligning the polarizing field with the *c*-axis restores a mirror symmetry, which protects a nodal ring state shown in Supplementary Fig. 19g. Thus, slight variations of temperature across the Neel transition and the application of a magnetic field are expected to expose an intricate magneto-topological phase diagram. The magneto-topological $Eu_3In_2As_4$ NWs and their rich phenomenology we predict here may thus provide a bedrock for a wealth of studies and applications. Among others, they may also serve as a superior platform for inducing topological superconductivity by interfacing them with a superconducting element. The intrinsic realization of chiral modes in them relieves the need to induce them by Zeeman as in semiconducting NWs. Accordingly, the topological energy gap that supports the topological modes in $Eu_3In_2As_4$ is more than two orders of magnitude larger than in semiconducting NWs (on the order of 10 meV relative to less than a 100 $\mu eV$, respectively).

Yet, beyond the particular intriguing material we have grown, we introduce here a new growth method of NWs based on vapor-solid topotactic mutual exchange in MBE. It will be exciting to explore other compounds that may benefit from this method and by this profoundly enrich the scope of materials that can be grown in the form of NWs. As initial candidates we suggest other

Zintl compounds from the orthorhombic *Pnnm* space group that share the topotactic requirement of common As substructure with WZ InAs. These include for instance the previously synthesized bulk compounds $X_3In_2Y_4$ (X=Ca/Sr/Ba and Y=As, P) on InY core. It is harder to predict which of these would support the mutual exchange process though it is natural to expect lighter elements would diffuse better through the pnictogen skeleton.

## Methods

### $Eu_3In_2As_4$ NW growth

Zintl $Eu_3In_2As_4$ grains were topotaxially grown on the surface of WZ InAs nanowire (NW). As a preliminary step, reclining and vertical WZ InAs NWs were grown by molecular beam epitaxy (MBE) on (001) and (111)B InAs substrates, respectively, using the gold-assisted vapor-liquid-solid (VLS) technique as described in the previous work[25,43–45]. The NWs surfaces were normally rounded which is typical of using $As_4$. To initiate the topotaxial cation exchange, the Eu shutter was opened while the In cell was cold and its shutter closed while the Arsenic flux was maintained. A 15-minute pause for adjusting the cells temperatures followed the InAs NWs growth. Right after opening of the Eu shutter the substrate temperature was ramped to a temperature in the range of 300 °C to 400 °C at a rate of 10 °C per minute. The topotaxial exchange growth was maintained for 1 to 4 hours for different samples. The temperature (and flux) of Eu and As were 450 °C ($3.5 \times 10^{-8}$ Torr) and 245 °C ($7.3 \times 10^{-6}$ Torr), respectively.

### Microscopy

The Zintl $Eu_3In_2As_4$ NWs were characterized by field emission scanning electron microscopy (FE-SEM, Zeiss Supra-55, 3 kV, working distance ~4 mm), transmission electron microscopy (TEM, Thermo Fisher Scientific Talos F200X, 200 kV). Energy dispersive spectroscopy (EDS) composition data and mapping images were obtained by TEM with an attached detector that is identical to the one in the scanning transmission electron microscopy (STEM).

High-resolution STEM images and analytical EDS maps were acquired in a double aberration-corrected Themis-Z microscope (Thermo Fisher Scientific Electron Microscopy Solutions, Hillsboro, USA, (TFS)) at an accelerating voltage of 200 kV. STEM images were recorded with a Fischione Model 3000 detector and a TFS BF detector. EDS hyperspectral data were obtained with a Super-X SDD detector and quantified with the Velox software (TFS) through background subtraction and spectrum deconvolution. STEM images were obtained with an electron probe with a convergence angle of 21 mrad and a primary beam current of less than 50 pA, the EDS maps were recorded at a beam current of 200 pA.

### Molecular dynamics simulations

To simulate the process of adding Eu atoms into the InAs NWs during the MBE growth, molecular dynamics calculations were performed. The simulations started with preparation of a [0001] oriented WZ NW core with six sidewalls. This NW contained 14 ×14 × 35 InAs bilayers. On both sides of the NW 1200 Eu atoms were arranged in 4 layers separated by 40 Å from the NW. The Tersoff type interatomic potentials[46] were applied to describe the interactions between In and As[47]. Since for Eu-In, Eu-As and Eu-Eu interactions such potentials are not available in literature, we assumed that between these three pairs of atoms the interactions are of Lennard-Jones (L-J) type. A standard L-J was used with the parameters $\varepsilon$=0.1, 0.4 and 0.05, and $\sigma$=2.6, 2.81 and 3.99 for Eu-In, Eu-As and Eu-Eu, respectively.

**SQUID**
We used scanning superconducting quantum interference device (SQUID) microscopy to search for magnetic signals from the NWs. A SQUID converts magnetic flux to voltage, allowing sensitive detection of magnetic fields[48]. The planar SQUID, in scanning configuration, allows mapping of the static magnetic landscape, and the local susceptibility[48]. The sensitive area of the SQUID used in this work, the pickup loop, has a diameter of 1.5 μm. Local susceptibility was measured using an on-chip coil to apply magnetic field (the field-coil, operated in this work at ~kHz, ~Gauss), while the pickup loop records the local response to the applied magnetic field. The positive signal in our data indicates a paramagnetic response. These measurements of susceptibility were plotted in units of $\Phi_0$ normalized by the sensitive area and the current in the field-coil.

We then performed these measurements as a function of temperature between 5K and 20K in order to characterize the magnetic behavior of the wires. We later plotted the data as the inverse of the change in susceptibility. The fit corresponds to the inverse of the Curie-Weiss law, yielding the Weiss constant.

In addition, the commercial SQUID magnetometer (MPMS3, Quantum Design) was also used to measure massive $Eu_3In_2As_4$ NWs. These NWs were dispersed on a $Si/SiO_2$ substrate (5×6 mm2), which was inserted into a straw and mounted in the MPMS3. The global magnetic measurement was carried out in the temperature range between 2K and 10K by varying the magnetic field from 500 Oe to 6,000 Oe.

**DFT**
Our first-principles calculations were performed with the Vienna Ab initio Simulation Package (VASP) using the projector-augmented wave method[49]. The Exchange-correlation functional based on generalized gradient approximation (GGA) parameterized by Perdew-Burke-Ernzerhof (PBE) was adopted[50]. To treat the correlation effect of localized 4f electrons of Eu, the DFT+U method by Dudarev et al. was employed with $U_{eff}$=7 eV[51]. The kinetic energy cutoff of the plane-wave basis was set to 400 eV. Brillouin zone integration was performed by using 7×3×11 Γ-centered k point mesh. We construct the maximally localized Wannier functions of In-5s orbital and As-4p orbitals by using the WANNIER90 package[52]. Surface state, Weyl nodes and Fermi arcs were investigated by using the Wannier Tools package[53].

**Data availability**

All data are available in the main text or the Supplementary Information.

**Acknowledgments**

With loving memory to our bright colleague and dear friend, Prof Perla Kacman (1948-2024). We deeply thank Michael Fourmansky and Israel Boterashvili for their professional assistance in the MBE growth of NWs, Dr. Markus Huecker in the MPMS measurements, Shai Rabkin and Xi Wang for additional scanning SQUID measurements. L.H. acknowledges the support of the Irving and Cherna Moskowitz Center for Nano and Bio-Imaging at the Weizmann Institute of Science. H.S. is an incumbent of the Henry and Gertrude F. Rothschild Research Fellow Chair. B.Y. acknowledges the financial support by the European Research Council (ERC Consolidator Grant "NonlinearTopo", No. 815869) and the ISF - Personal Research Grant (No. 2932/21). N.R., B.K. were supported by European Research Council Grant (No. ERC-2019-COG-866236) and Israeli Science Foundation Grant (No. ISF-228/22). P.K. acknowledges the scientific cooperation project financed by the Polish Academy of Science and the Israel Academy of Science and Humanities. M.Z.-K. was supported by The National Center for Research and Development grant (No. EIG CONCERT-JAPAN/9/56/AtLv-AIGaN/2023). H.B. and H.S. acknowledge support from the European Research Council (ERC-PoC TopoTapered – 101067680) and the Israel Science Foundation (Grant 1152/23).

**Author contributions**

M.S.S., H.S., H.B. conceptualized this research. H.S. and M.S.S. designed and carried out the MBE growths. M.S.S., L.H. did the electron microscopy imaging with EDS analysis and modeling of the interfacial crystal structure. Y.Z., B.Y. performed DFT calculations and analyzed the topological states. N.R., B.K. performed scanning SQUID measurements and analyzed the data. M.Z.-K., P.K. provided results of the molecular dynamics calculations. L.H., B.Y., B.K., H.S., H.B. wrote the manuscript. M.S.S., L.H., H.S., H.B. prepared the figures for the manuscript and Supporting Information. All authors contributed to the discussion and manuscript preparation.

**Competing interests**

The authors declare a competing interest related to a patent application. M.S.S., L.H., H.S. and H.B. are inventors on US provisional No. 63/472,598. The patent was filed on June 13, 2023, under YEDA RESEARCH AND DEVELOPMENT CO. LTD.

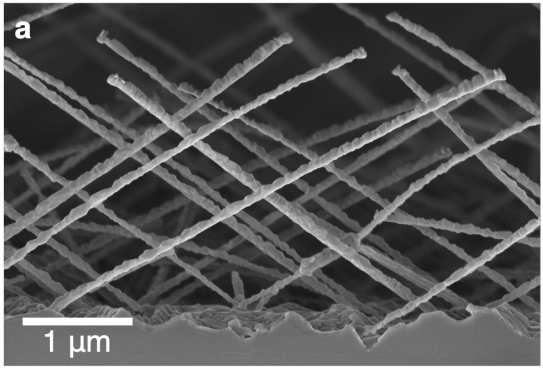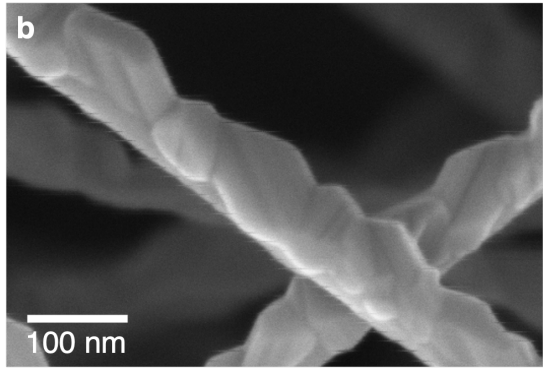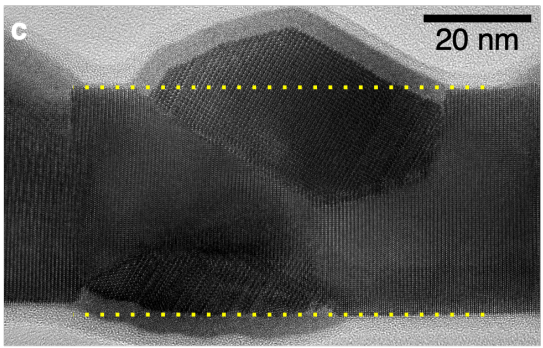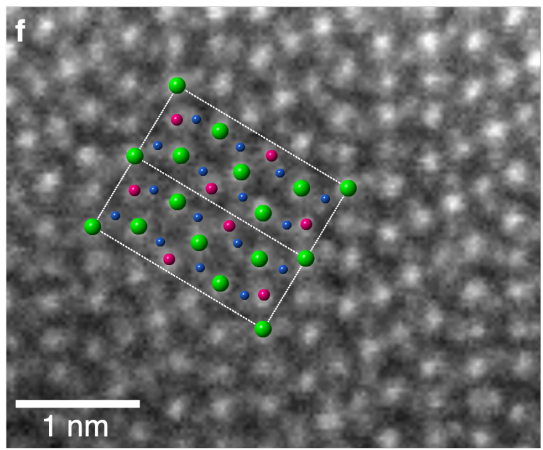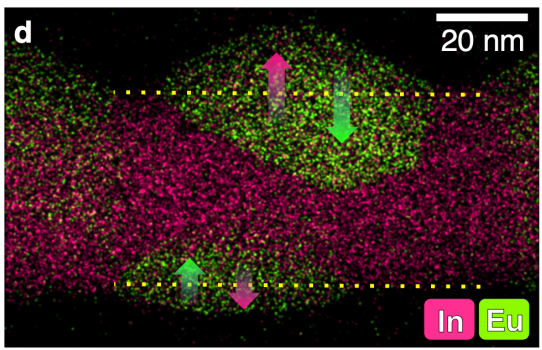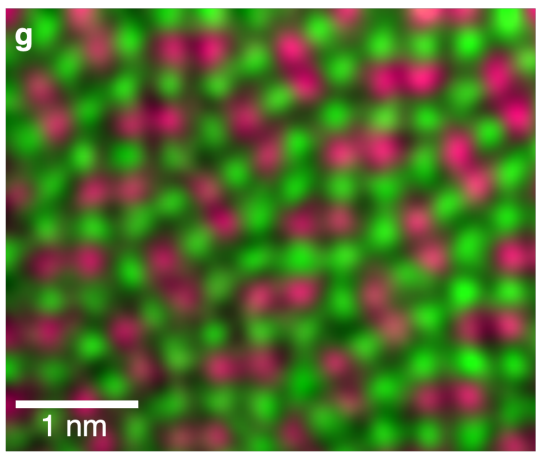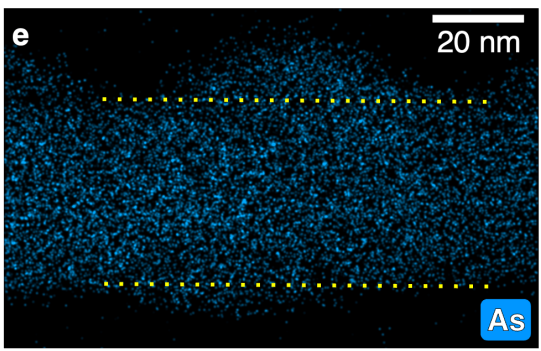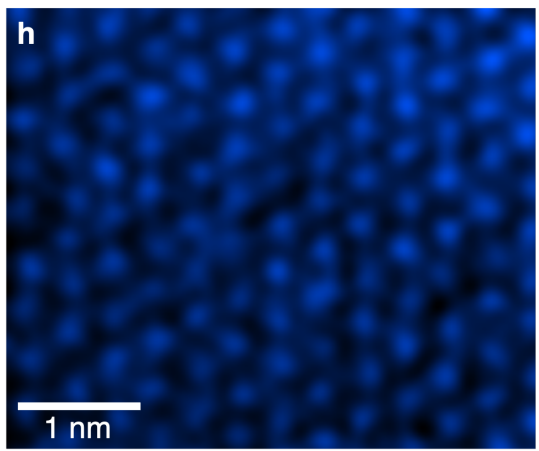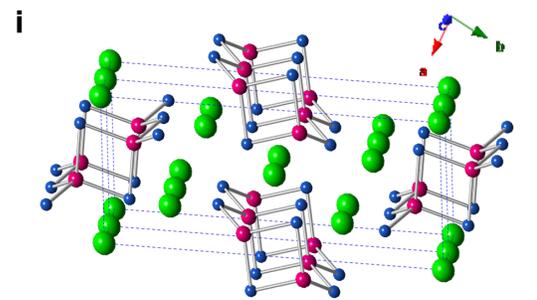

**Fig. 1. Exchange growth of Zintl $Eu_3In_2As_4$ shell on WZ InAs core.** **(a)** Cross-section-viewed SEM image of $Eu_3In_2As_4$ shell grown on WZ InAs core NWs pregrown on (001) InAs substrate. **(b)** A magnified SEM image of $Eu_3In_2As_4$ NW revealing two types of grain with clear facets. **(c)** TEM image of $Eu_3In_2As_4$ crystallites on WZ InAs core. The dotted lines sketch the boundaries of the WZ InAs core before the shell was deposited. **(d)** EDS maps of Eu and In (green and magenta, respectively) showing their uniform distribution on either side of the core boundary. **(e)** EDS map of As showing its uniform distribution across the core and the shell. **(f)** High resolution STEM-HAADF image showing the crystalline structure of $Eu_3In_2As_4$ along the <001> zone axis. 2 unit cells are superimposed. **(g, h)** EDS map of Eu, In and As atoms in green, magenta and blue, respectively. **(i)** The crystal structure of Zintl $Eu_3In_2As_4$.

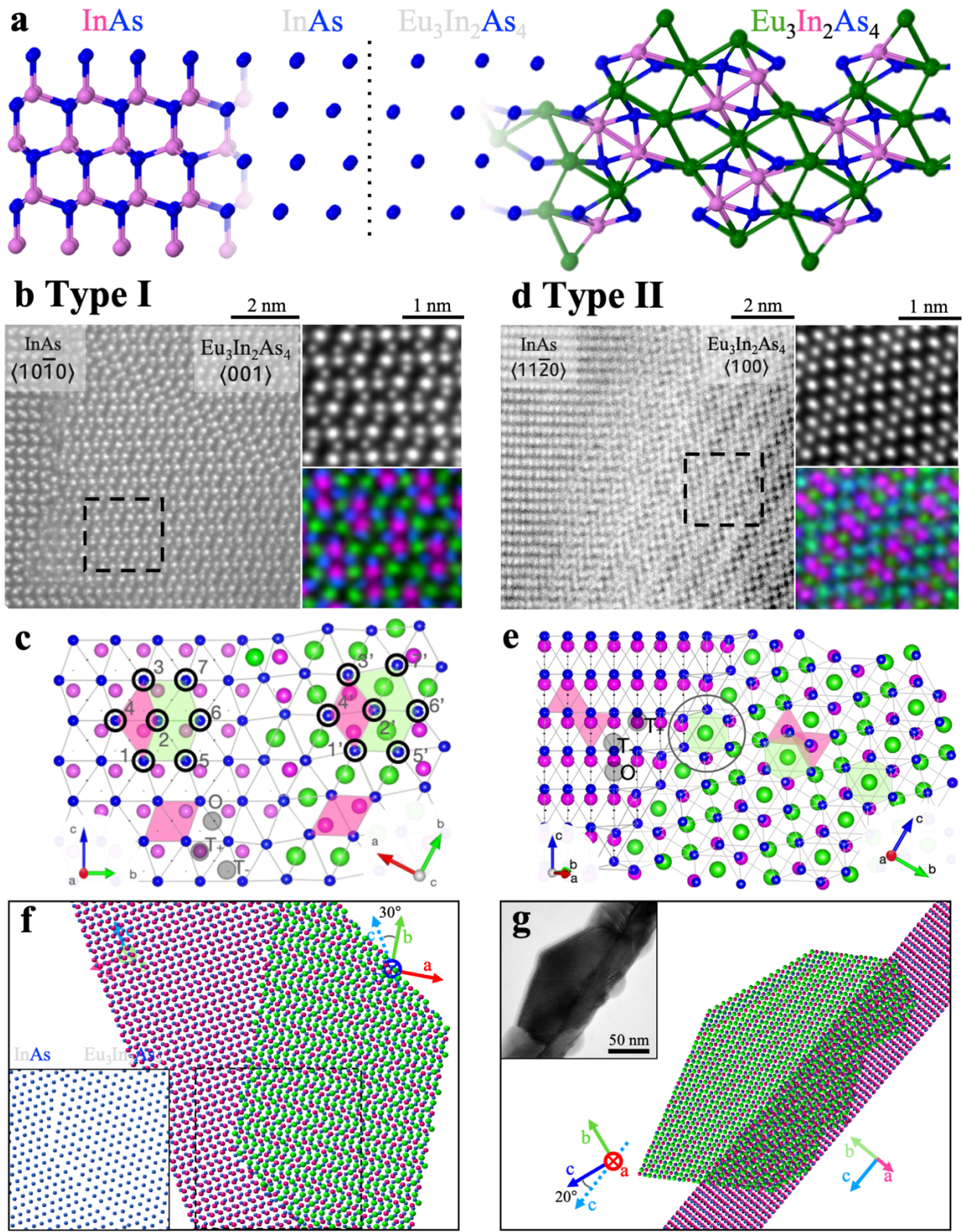

**Fig. 2. Topotaxial transformation of hexagonal InAs into orthorhombic Eu₃In₂As₄.** (a) Crystal structure of WZ InAs next to Eu₃In₂As₄ and their corresponding As sub-structure in between. (b, d) Atomically resolved STEM images of the transformation interfaces between InAs and

Eu$_3$In$_2$As$_4$ for the two orientation relationships type I - $\langle 10\bar{1}0\rangle_{InAs} \parallel \langle 001\rangle_{EuInAs}$ and $(01\bar{1}0)_{InAs} \parallel (\bar{2}30)_{EuInAs}$; type II - $\langle 1\bar{2}10\rangle_{InAs} \parallel \langle 100\rangle_{EuInAs}$ and $(20\bar{2}1)_{InAs} \parallel (\bar{1}40)_{EuInAs}$. The top right panels show the magnified Eu$_3$In$_2$As$_4$ that matches the *a*- and *c*-axis projection of the orthorhombic phase. The precise location of Eu, In, and As is resolved in the atomic-scale EDS map underneath (green, blue, and pink, respectively) in the bottom right panels. **(c, e)** Atomistic model for the respective viewing direction. The edges of the polyhedral network spanned by the As sub-lattices are outlined. *T+*, *T-* and *O* mark tetrahedral and octahedral coordination sites in InAs. Pairs of tetrahedra (red filling) around In in the Eu$_3$In$_2$As$_4$ form the polyanionic InAs chains that align along the *c*-axis, surrounded by octahedrally coordinated Eu (green filling). For type I an exemplary mapping of As sites 1-7 in InAs to As sites 1'-7' in Eu$_3$In$_2$As$_4$ is depicted. Interface type II involves an interfacial octahedral defect plane that leads to a symmetry flip (circular marker). **(f)** Modeling of the atomic crystal structure of type I (from Fig. 1c). The configuration with a ~30° angle between the *b*-axis of the grain and the *c*-axis of the core allows the As-substructure to continue seamlessly across the interface. (The inset corresponds to the dotted square without Eu and In atoms.) **(g)** Modeling of the type II crystal structure built from the inset of the TEM image. This model depicts the only center grain (the other is omitted). A part of the area is overlapped by two structures at a ~20° angle between two *c*-axes. The HRTEM images are shown in Supplementary Fig. 5.

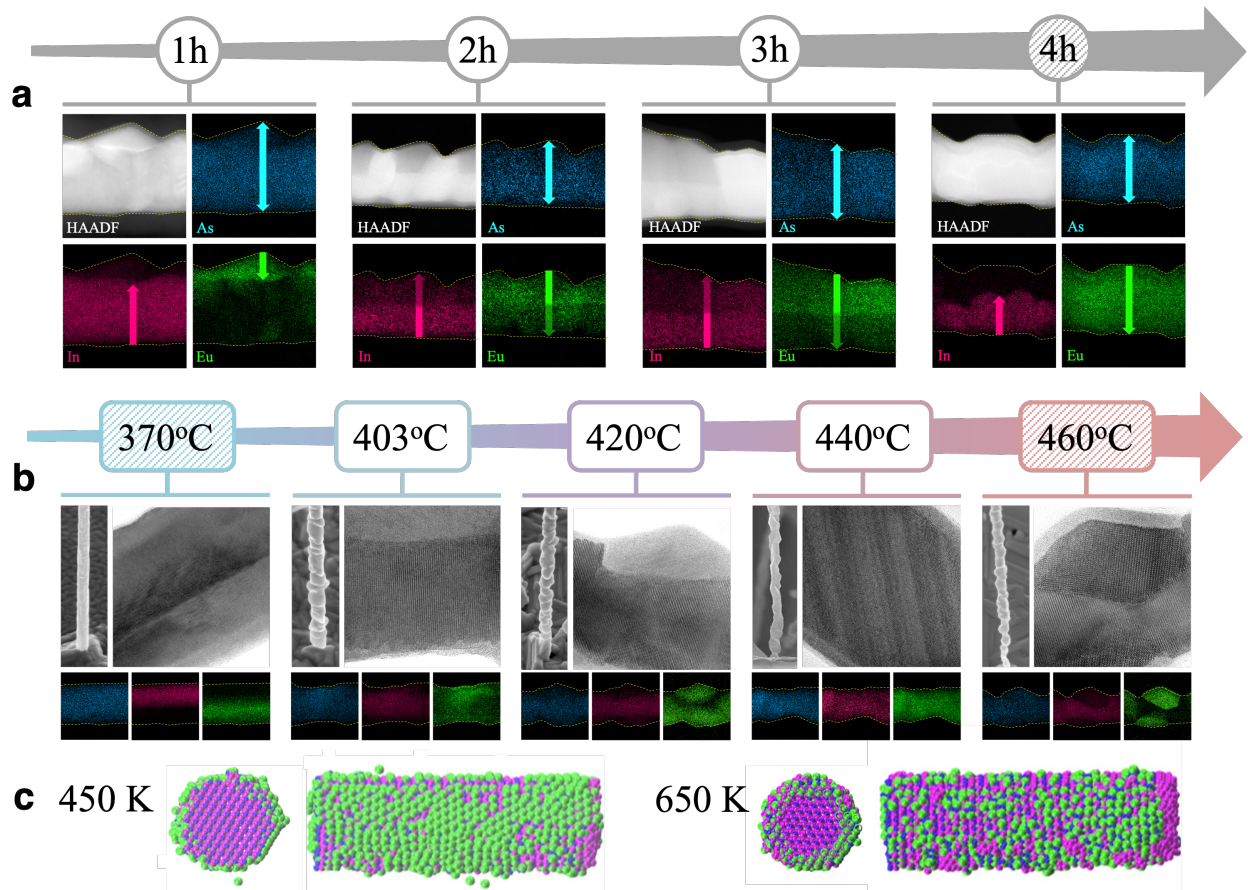

**Fig. 3. Parameter space for topotactic mutual-exchange growth of Zintl Eu$_3$In$_2$As$_4$ NWs. (a)** Growth time - increasing the Eu-As deposition time from 1 to 4 hours (left to right) under the same substrate temperature of 420 °C the interface between the Eu$_3$In$_2$As$_4$ shell and InAs core progresses until at 4 hours the 50 nm InAs core is fully consumed and a Eu-As shell begins to grow. **(b)** Substrate temperature - increasing the substrate temperature from 370 °C to 460 °C (left to right) under the same growth time of 2 hours. At 370 °C exchange does not occur and a Eu-As shell grows on the InAs core. Mutual exchange commences at slightly higher temperatures and becomes more efficient with increasing temperature. At 440 °C the core is fully consumed after 2 hours, and at 460 °C a new growth dynamics appears resulting in a segregation of Eu$_3$In$_2$As$_4$ crystallites among InAs segments. **(c)** Molecular dynamics simulation of depositing Eu-like atoms on WZ InAs-like NW at two different temperatures. At 450 K (left) the Eu-like atoms form a shell, while at 650 K (right) they mutually-exchange with In atoms from the core.

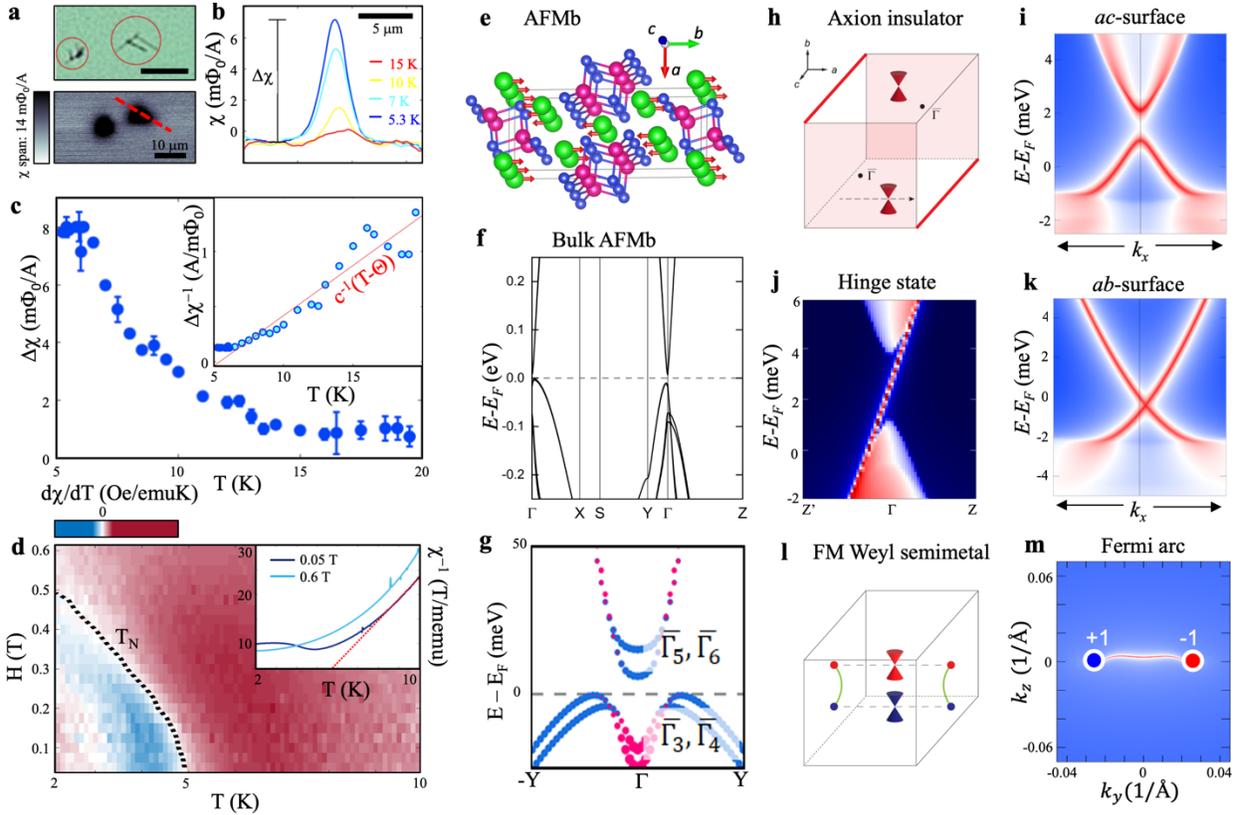

**Fig. 4. Magnetic and electronic characterization of Zintl Eu$_3$In$_2$As$_4$.** (a) Optical and SQUID AC susceptibility mapping (top and bottom panels, respectively) of Eu$_3$In$_2$As$_4$ NWs. (b) AC susceptibility measured across the NWs (along dashed line in A) at various temperatures. (c) AC susceptibility of the NWs versus temperature showing change of trend at about 6.5 K. The inset shows the inverse susceptibility displaying deviation from a Curie-Weiss law (red line) below 6.5 K. (d) Plot of dχ/dT(T, H) where the Neel transition appears as the line of zero slope (maxima in χ(T)). The inset shows inverse susceptibility at low and finite magnetic fields and a fit to Curie-Weiss law. (e) Crystal structure of Eu$_3$In$_2$As$_4$ with a C-type AFMb order. (f) Electronic band structure of AFMb phase with spin-orbit coupling obtained by GGA+$U$ calculations. (g) Orbital texture at the vicinity of the gap showing gap inversion within the AFMb state. (h) Schematic of AFM axion insulator state. Gapless surface and hinge states are marked as red while *ac*-plane and *bc*-plane are gapped. The dashed arrow line depicts the location of the unpinned Dirac cone and *k*-path in (i). (i) Surface states of AFM phase at *ab*-plane ($k_y = 5 \times 10^{-5}$/Å). (j) Numerical calculated hinge state. (k) Surface states of AFM phase at *ac*-plane ($k_y = 0$/Å). (l) Schematic of FM Weyl semimetal phase. Fermi arcs (green) on the surfaces connecting the projections of two Weyl nodes with opposite chirality (red and blue). They are distributed along Y'-Γ-Y in FMb phase. (m) Fermi arcs at *bc*-plane.

# Supplementary Information

**Topotaxial Mutual-Exchange Growth of Magnetic Zintl $Eu_3In_2As_4$ Nanowires with Axion Insulator Classification**

Man Suk Song[1], Lothar Houben[2], Yufei Zhao[1], Hyeonhu Bae[1], Nadav Rothem[3], Ambikesh Gupta[1], Binghai Yan[1], Beena Kalisky[3], Magdalena Zaluska-Kotur[4], Perla Kacman[4], Hadas Shtrikman[1]\*, Haim Beidenkopf[1]\*

[1]Department of Condensed Matter Physics, Weizmann Institute of Science; Rehovot, 7610001, Israel.
[2]Department of Chemical Research Support, Weizmann Institute of Science; Rehovot, 7610001, Israel.
[3]Department of Physics and Institute of Nanotechnology and Advanced Materials, Bar-Ilan University; Ramat Gan, 5290002, Israel.
[4]Institute of Physics, Polish Academy of Sciences; Warsaw, PL-02-668, Poland.

\*Corresponding author: hadas.shtrikman@weizmann.ac.il (H.S.); haim.beidenkopf@weizmann.ac.il (H.B.)

**$Eu_3In_2As_4$ growth and analysis**

In this section we add more information gathered during our intensive study of the formation of $Eu_3In_2As_4$ by the evaporation of Eu and As on pre-grown WZ InAs nanowires (NWs). We provide additional EDS data, which confirm the composition of the $Eu_3In_2As_4$. This data is presented in Supplementary Figs. 1 and 2.

We also offer more structural and compositional information on the $Eu_3In_2As_4$ type II grains which is provided in Supplementary Figs. 3, 4 and 5. Finally additional information is provided in Supplementary Figs. 6 and 7 which relate to the prominent effect of the temperature gradient along the nanowires which strongly affects the mutual exchange and thus the distribution of type I and type II grains along the NWs.

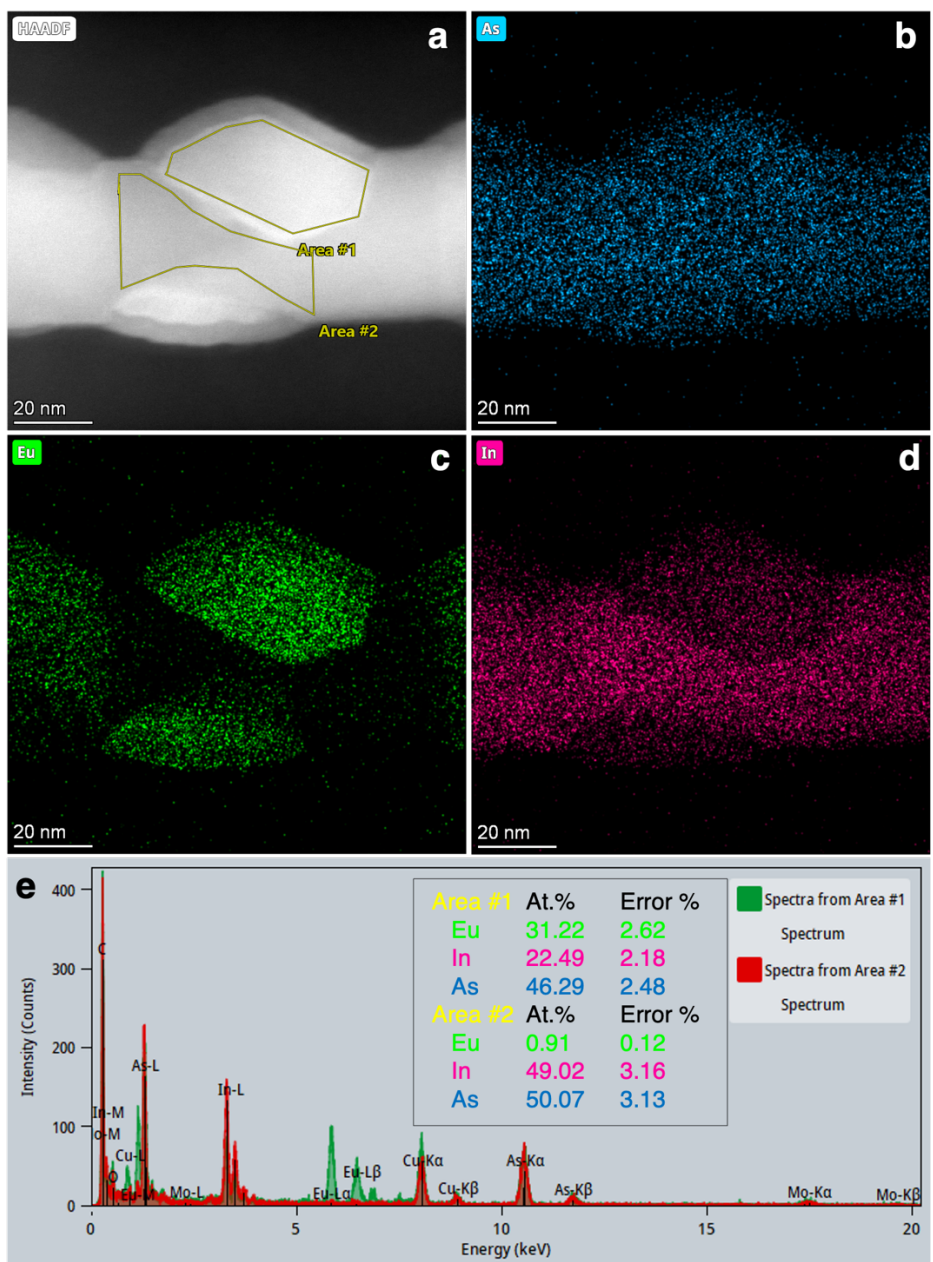

**Supplementary Fig. 1. Elemental composition of a Type I Eu$_3$In$_2$As$_4$ grain and WZ InAs core. (a)** The Eu$_3$In$_2$As$_4$ and InAs areas were selected in a HAADF image for EDS measurement. **(b, c, d)** EDS elemental maps showing the distribution of As, Eu, and In. **(e)** Cumulated EDS spectra extracted from the two areas, Area #1 and Area #2, in (a). The inset shows quantification of the spectra.

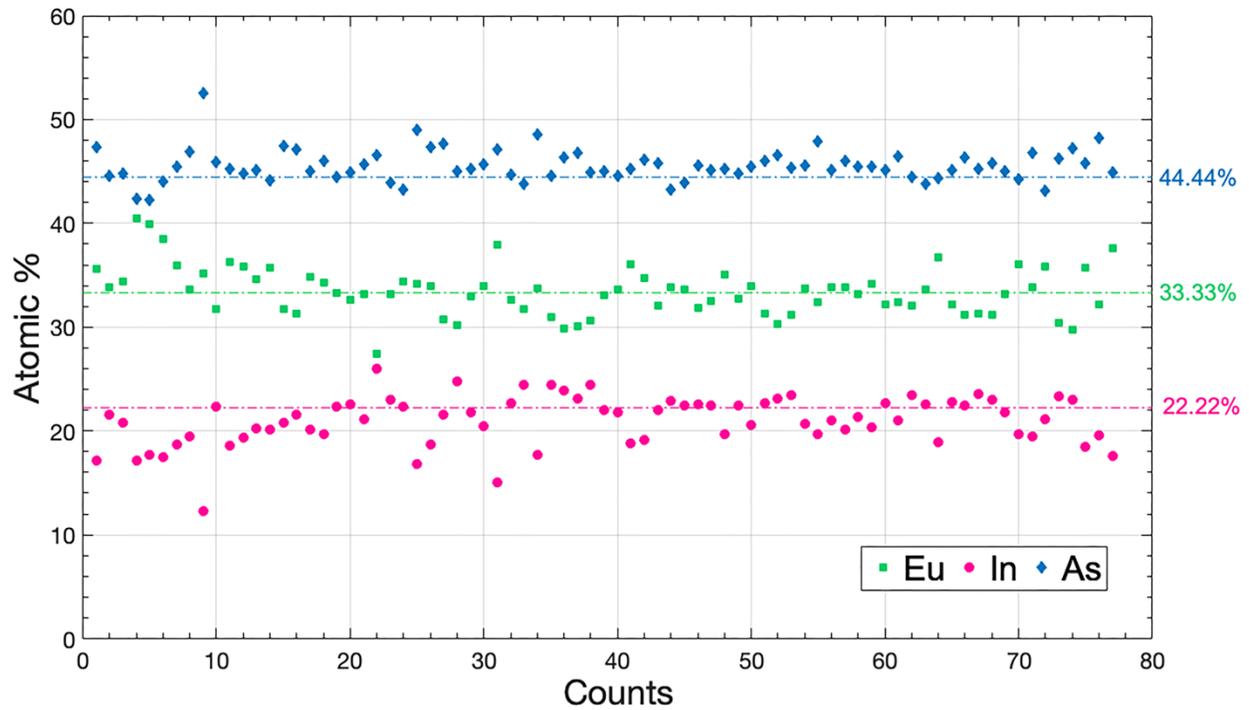

**Supplementary Fig. 2.** EDS atomic percentage distribution of Eu (green square), In (pink circle), and As (blue diamond) for selected $Eu_3In_2As_4$ grains was measured by STEM-EDS. Three dash-dotted guidelines signify an ideal composition of $Eu_3In_2As_4$.

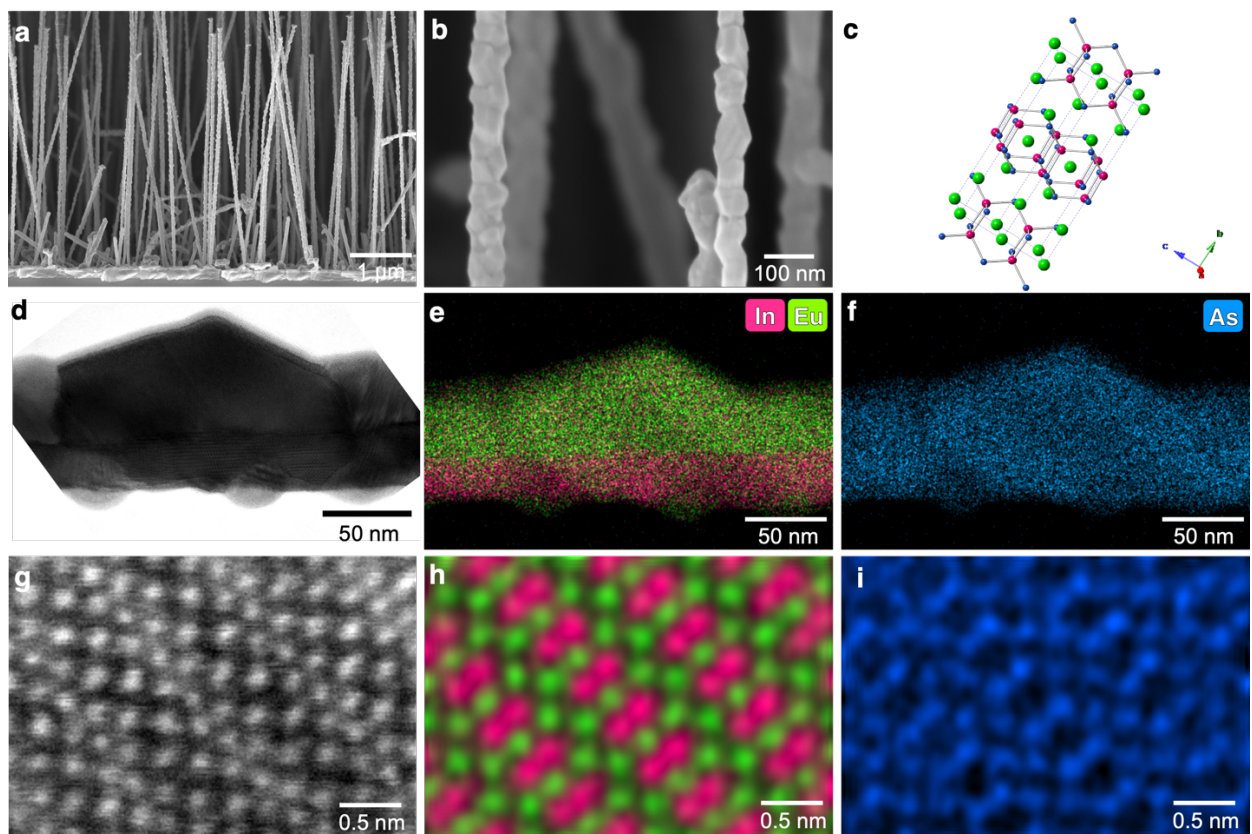

**Supplementary Fig. 3. (a)** Cross-section-viewed SEM image of as-grown vertical $Eu_3In_2As_4$ NWs on a InAs (111)B substrate. (**b**) A magnified SEM image of $Eu_3In_2As_4$ NW revealing two types of grain with clear facets. **(c)** 2×1 unit cells of $Eu_3In_2As_4$. Eu, In, and As atoms are represented as green, pink, and blue balls, respectively. **(d)** HRTEM image of type II $Eu_3In_2As_4$ crystallites on WZ InAs core. The other magnified HRTEM ones are in the following Supplementary Fig. 4. **(e, f)** EDS maps of In, Eu, As elements corresponding to (d). **(g)** Atomic resolved STEM-HAADF image evidently shows crystalline structure of $Eu_3In_2As_4$ along the <100> zone axis. **(h, i)** The corresponding EDS elemental maps from (g).

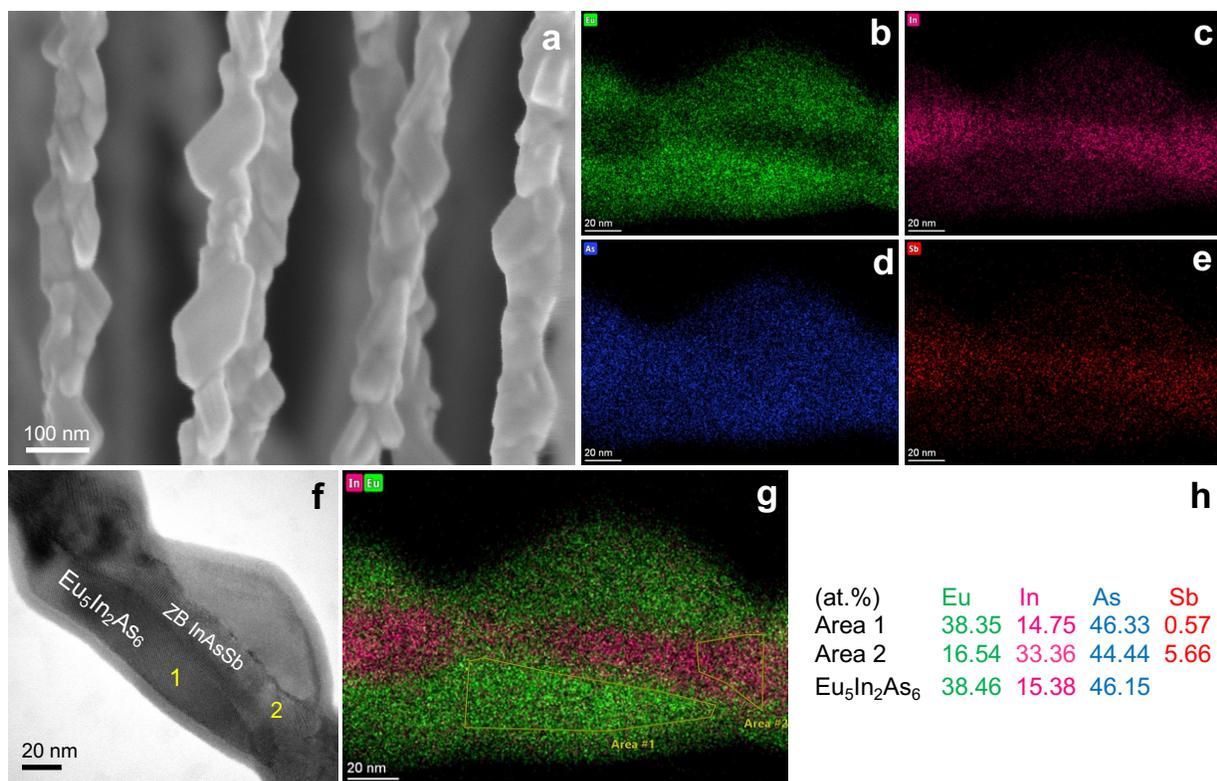

**Supplementary Figure 4:** (a) SEM of as grown $Eu_5In_2As_6$ NWs formed by exchange reaction between EuAs and the pre-grown $InAs_{1-x}Sb_x$ (x=0.05) zinc-blende core. The low concentration of Sb enforces a ZB structure[1,2] but does not affect the topotactic exchange process. (b–e) EDS mapping of Eu, In, As and Sb. (f) TEM of a segment of the $Eu_5In_2As_6$ NW. (g) EDS mapping of In and As indicating two mapped areas 1 and 2 which indicate that the composition of area 1 refers to $Eu_5In_2As_6$ as indicated in (h).

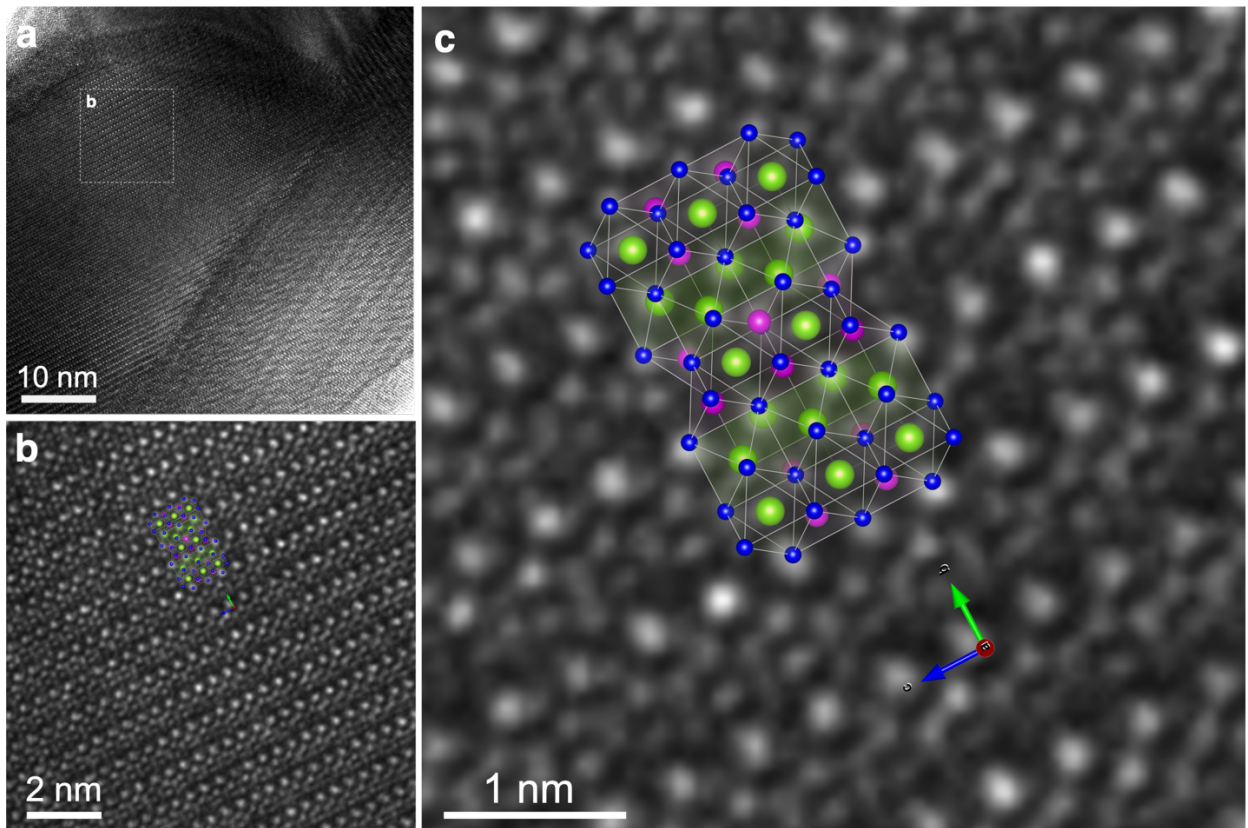

**Supplementary Fig. 5.** **(a)** An enlarged HRTEM image from the inset of Fig. 2g as well as Supplementary Fig. 3d. This shows both clear single crystalline Type II Eu$_3$In$_2$As$_4$ (upper left) and WZ InAs (lower right) with moiré fringes caused by two overlapped structures. **(b, c)** Further enlarged HRTEM images from (a). The 1×2-unit cells of Eu$_3$In$_2$As$_4$ are overlaid on HRTEM images with axial vectors along the <100> zone axis.

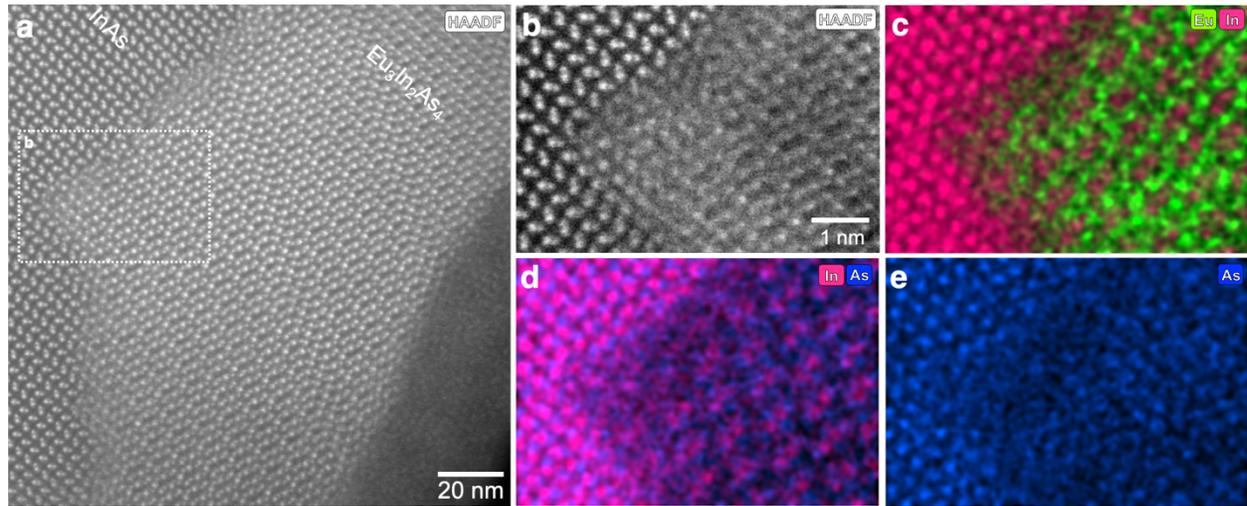

**Supplementary Fig. 6. (a)** STEM-HAADF image of phase boundary between InAs and Type I $Eu_3In_2As_4$ in the $\langle 10\bar{1}0 \rangle_{InAs}$ and $\langle 001 \rangle_{Eu_3In_2As_4}$ zone axes. **(b)** Enlarged HAADF image and EDS elemental maps of **(c)** Eu and In, **(d)** In and As, and **(e)** As in the interfacial region marked by a dotted line rectangle. It is noteworthy that the As sub-structure is maintained comprising a continuous framework regardless of phase in (e).

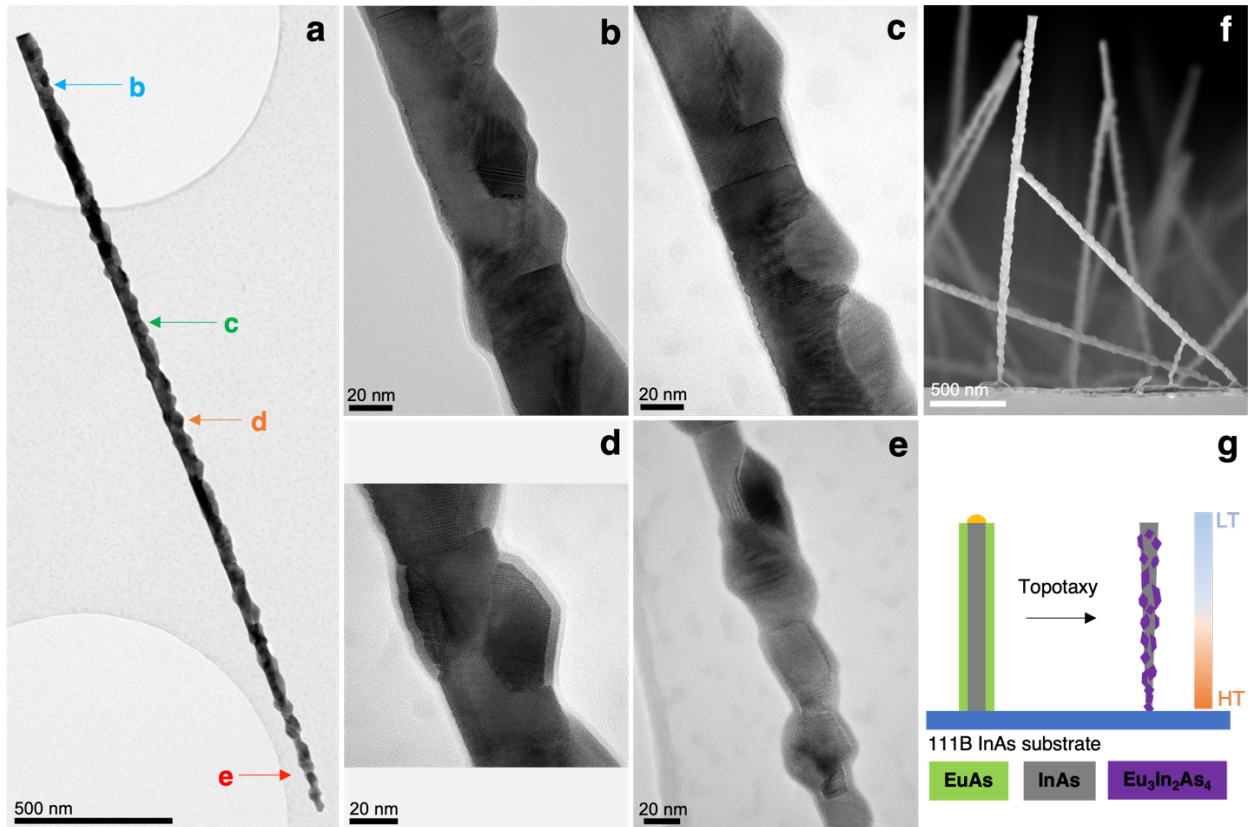

**Supplementary Fig. 7. The Eu$_3$In$_2$As$_4$ NW tapered along the extent of the exchange growth process that depends on the temperature. (a)** A TEM image of the as-grown NW at 460 °C. **(b–e)** Enlarged TEM images marked by arrows in (a). Such NWs easily collapse due to a narrow base of the NW in the SEM image **(f)**. **(g)** A schematic illustrating that the reverse tapering of the Eu$_3$In$_2$As$_4$ NW is attributed to the temperature gradient near the substrate during topotactic exchange growth.

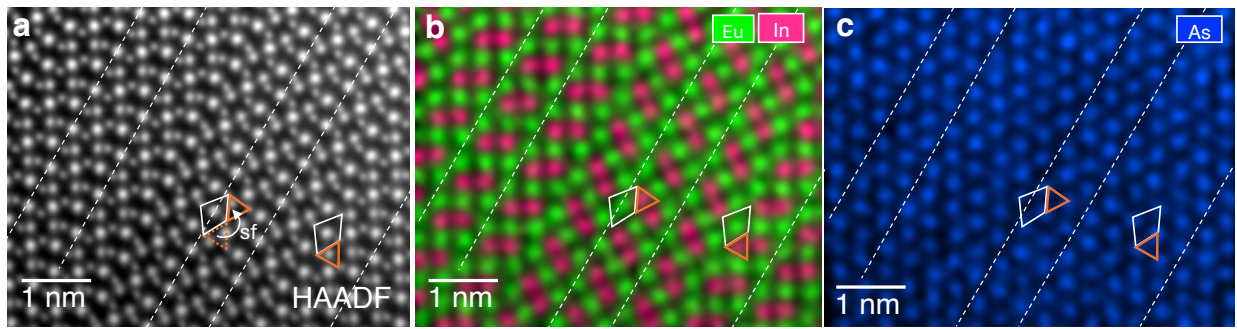

**Supplementary Fig. 8. Stacking faults in type I Eu$_3$In$_2$As$_4$ grain viewed along the <001> zone axis. (a)** Atomically resolved dark-field STEM image. **(b, c)** Corresponding EDS elemental maps of In, Eu and As atoms in green, magenta and blue, respectively. Dashed lines mark the stacking faults (sf) on (010) planes of Eu$_3$In$_2$As$_4$. The stacking fault occurs when In occupies tetrahedral sites (orange triangle) on the opposite side of Eu octahedral coordination sites (white parallelogram) along the (100) direction. The hexagonal As lattice is continuous, but it shows a small rigid body translation around the stacking fault.

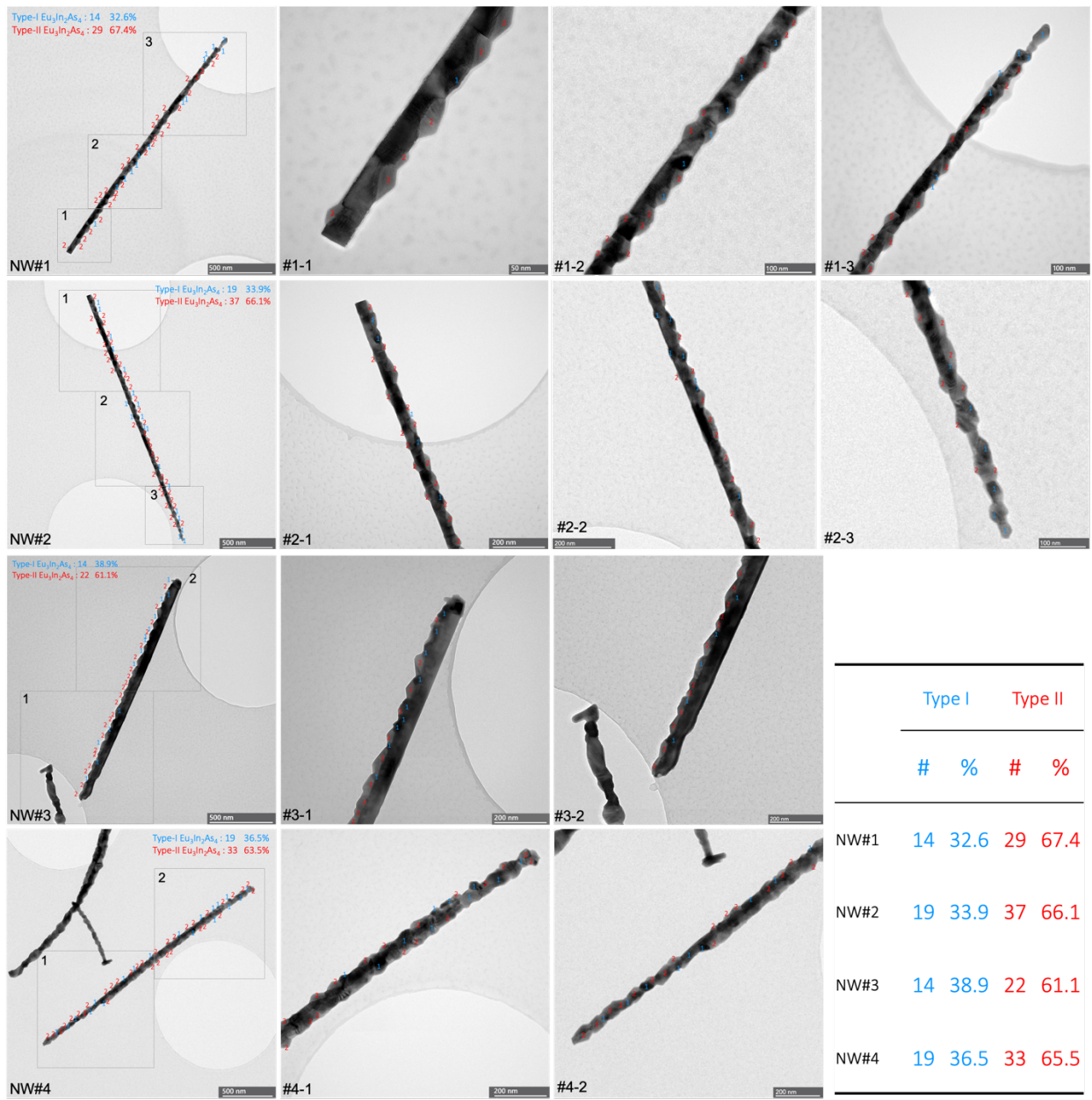

**Supplementary Fig. 9.** TEM images used to demonstrate which type of $Eu_3In_2As_4$ is more prominent along the NW. The higher-magnitude TEM images were used to determine whether a grain is Type I (blue) or II (red). Type I and II grains were randomly distributed, and the number of Type II is approximately double that of Type I.

**Modeling of Eu atoms evaporated on WZ InAs NWs by molecular dynamics simulations**

Molecular dynamics simulations were performed at two different temperatures, 650K, using the Lammps package for $10^6$ steps. The results of the simulation are shown in Fig. 3c. As can be seen in the NW cross-section view in Fig. 3c, at the lower temperature Eu atoms reaching the NW stay at their side surfaces (the left side), while at the higher temperature they penetrate into the interior of the crystal (the right side). In the side-view of Fig. 3c we see that at 450K the Eu atoms form a regularly ordered layer on the NW surface. At the higher temperature, many Eu atoms disappear from this layer and enter into the crystal. Interestingly, a further increase in temperature does not change the result – even a temperature as high as 900K leads to a similar result as 650K in terms of the penetration of Eu atoms into the NW. The only difference is that in such high temperatures, the NW side facets became very disordered. Similar effects have been also obtained in simulations in which a NW with Eu atoms deposited on the surface is heated. In this case after some time the Eu atoms also start to enter the interior of the NW and the final picture depends only on the annealing temperature.

Another molecular dynamics process of simulation was performed by $10^6$ steps at 700 K, for In and As atoms with Tersoff type potential and L-J potential for other pairs of atoms. The L-J potential is similar to the one used in our previous work[3] with slightly modified range of repulsive part of interaction. In Supplementary Fig. 8a, the initial core NW consists of three segments, 12 bilayers each: one ZB oriented in the [111] direction (the left side) and the other two are WZ NWs oriented in the [0001] direction. WZ1 has six side walls while WZ2 twelve (six $\{1\bar{1}00\}$ and six $\{11\bar{2}0\}$) side walls.

The results of the simulations are shown in Supplementary Fig. 8b. On ZB surface one can see V shape lines of Eu atoms, signifying a typical mosaic structure. In contrast, in the WZ parts parallel lines of Eu atoms (green) appear on the surface. Here Eu atoms penetrate deep into the NW while at the same time In atoms (magenta) go up to the surface, on which a new crystal structure starts to be built from initially glassy arrangement of atoms. Supplementary Figures 8c and 8d show the potential profiles of Eu (green), In (magenta), and As (blue) atoms.

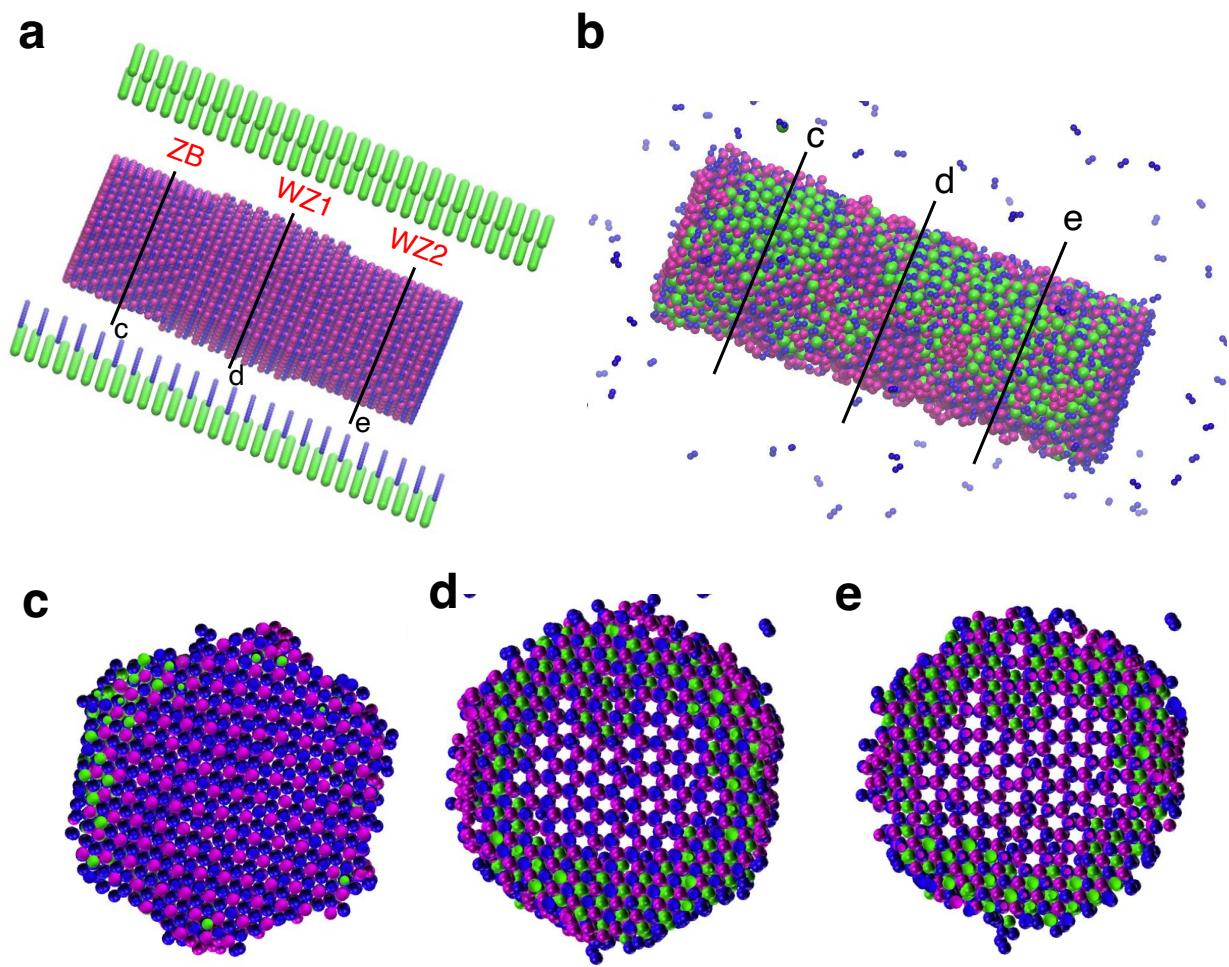

**Supplementary Fig. 10. (a)** Initial core InAs NW with ZB part in the [111] direction and two types of WZ segments in the [0001] direction, 12 bilayers each. WZ1 has six $\{11\bar{2}0\}$ side walls while WZ2 twelve (six $\{1\bar{1}00\}$ and six $\{11\bar{2}0\}$) side walls. On both sides of the NW 4 layers (three layers of Eu and one layer of As) were arranged, as a source of freely moving adatoms. Eu atoms are green while In and As atoms are magenta and blue, respectively. **(b)** In the obtained simulations the ZB surface exhibits a mosaic pattern with convex V-shaped lines of Eu atoms. The WZ sections present parallel lines of Eu atoms, which can deeply penetrate into the NW. **(c-e)** Cross-sections of a ZB and two WZ locations along the NW (marked in b with lines). The main difference between the zincblende and wurtzite segments can be appreciated by the level of Eu penetration into the NW core. Under the same conditions, a few Eu atoms penetrate into the ZB core in a scattered way, while in the WZ segments a uniform and ordered shell forms.

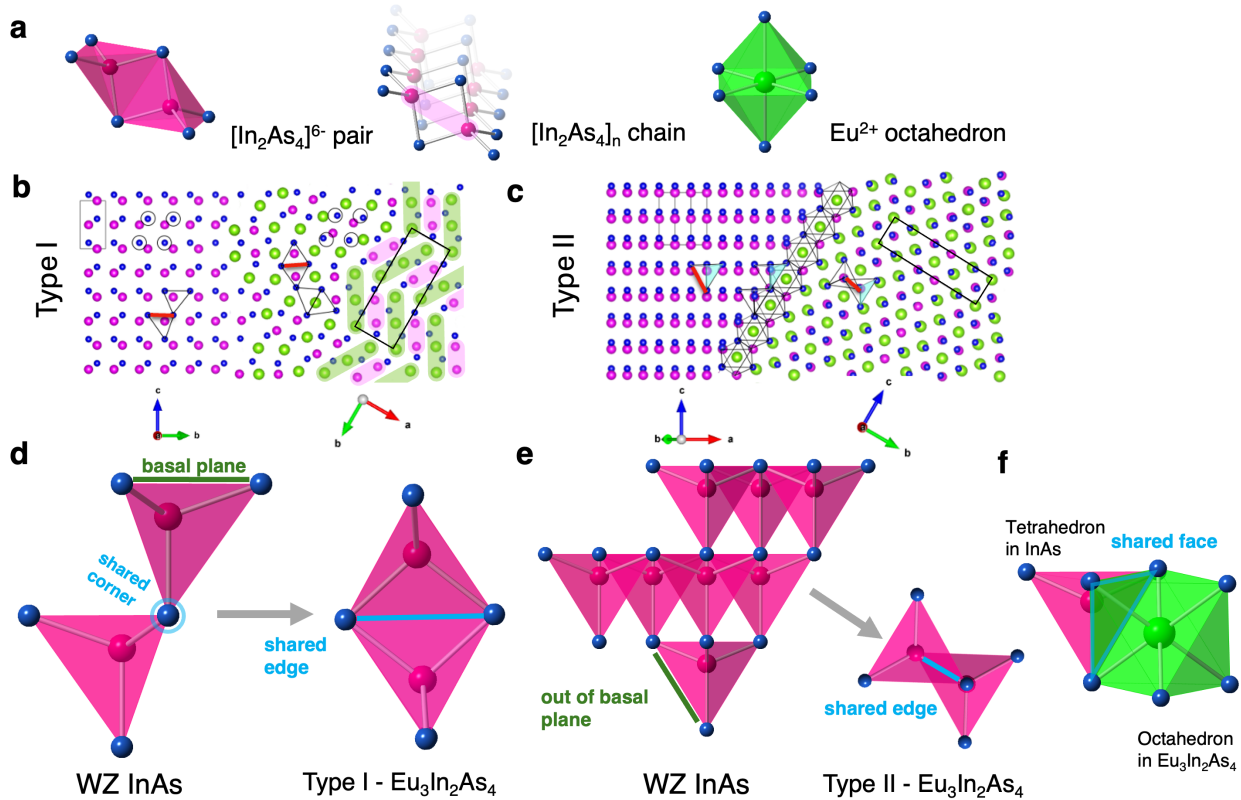

**Supplementary Fig. 11.** (a) Illustration of $[In_2As_4]^{6-}$ pair (left) and $[In_2As_4]_n$ chain , and $Eu^{2+}$ octahedron (right) (right) in the Zintl phase $Eu_3In_2As_4$. They consist of two distorted InAs tetrahedra combined by means of a shared edge. Two indium atoms share two arsenic ones inside, and four arsenic atoms outside share other tetrahedra pairs so that a $[In_2As_4]_n$ chain forms along the c-axis of $Eu_3In_2As_4$. **(b, c)** The atomic models of Type I and II like Fig. 2c and e in the main text, respectively. **(d)** In type I, one basal plane tetrahedral edge in corner-shared WZ InAs turns to a shared edge in one of the rotations of the $[In_2As_4]_n$ chains. **(e)** In type II, on the other hand, only out-of-basal plane tetrahedra edges in InAs become shared edges in the $[In_2As_4]_n$ chains. **(f)** In addition, the shared faces exist in the interface between InAs and $Eu_3In_2As_4$ as mentioned in Fig. 2e of the main text.

**Scanning SQUID measurements**

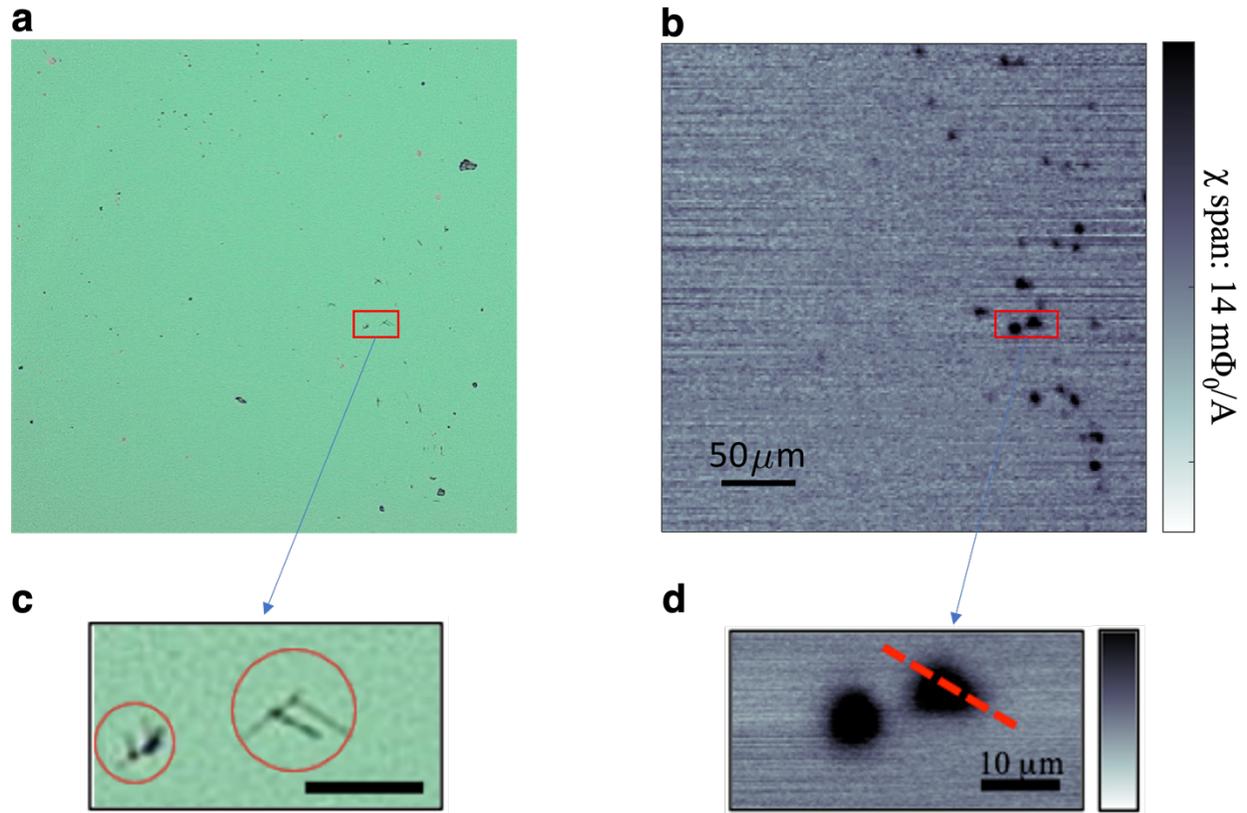

**Supplementary Fig. 12.** (a) Optical microscopy image of $Eu_3In_2As_4$ nanowires on a $SiO_2$ chip. (b) AC susceptibility map of the region shown in (a), taken at 5K. (c) Close-up of the region by a red rectangle in (a). (d) High-resolution AC susceptibility map of the region indicated by the red rectangle in (b).

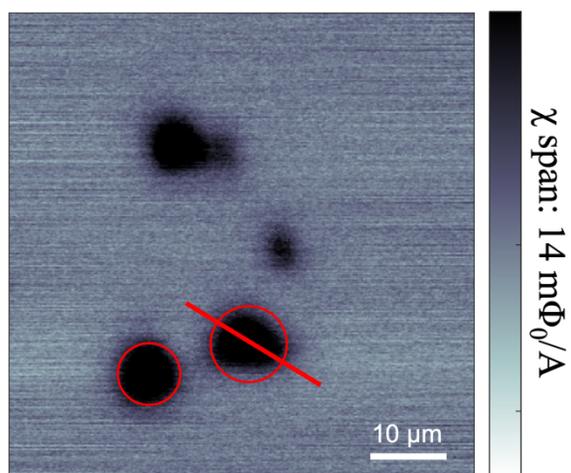 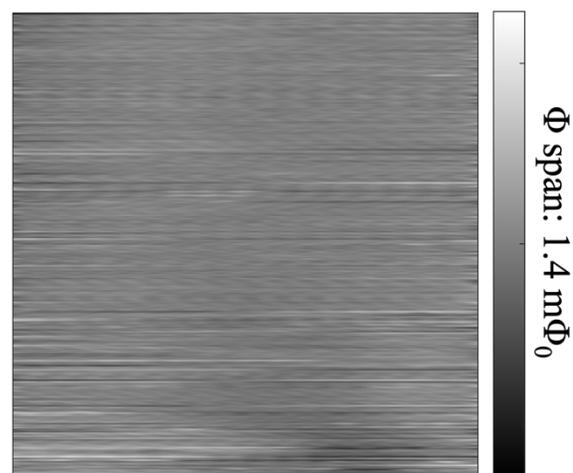

**Supplementary Fig. 13.** Comparison of the high-resolution **(a)** AC susceptibility and **(b)** DC magnetism maps of the region near the red rectangle in Supplementary Fig. 10b. No DC magnetization signal was detected.

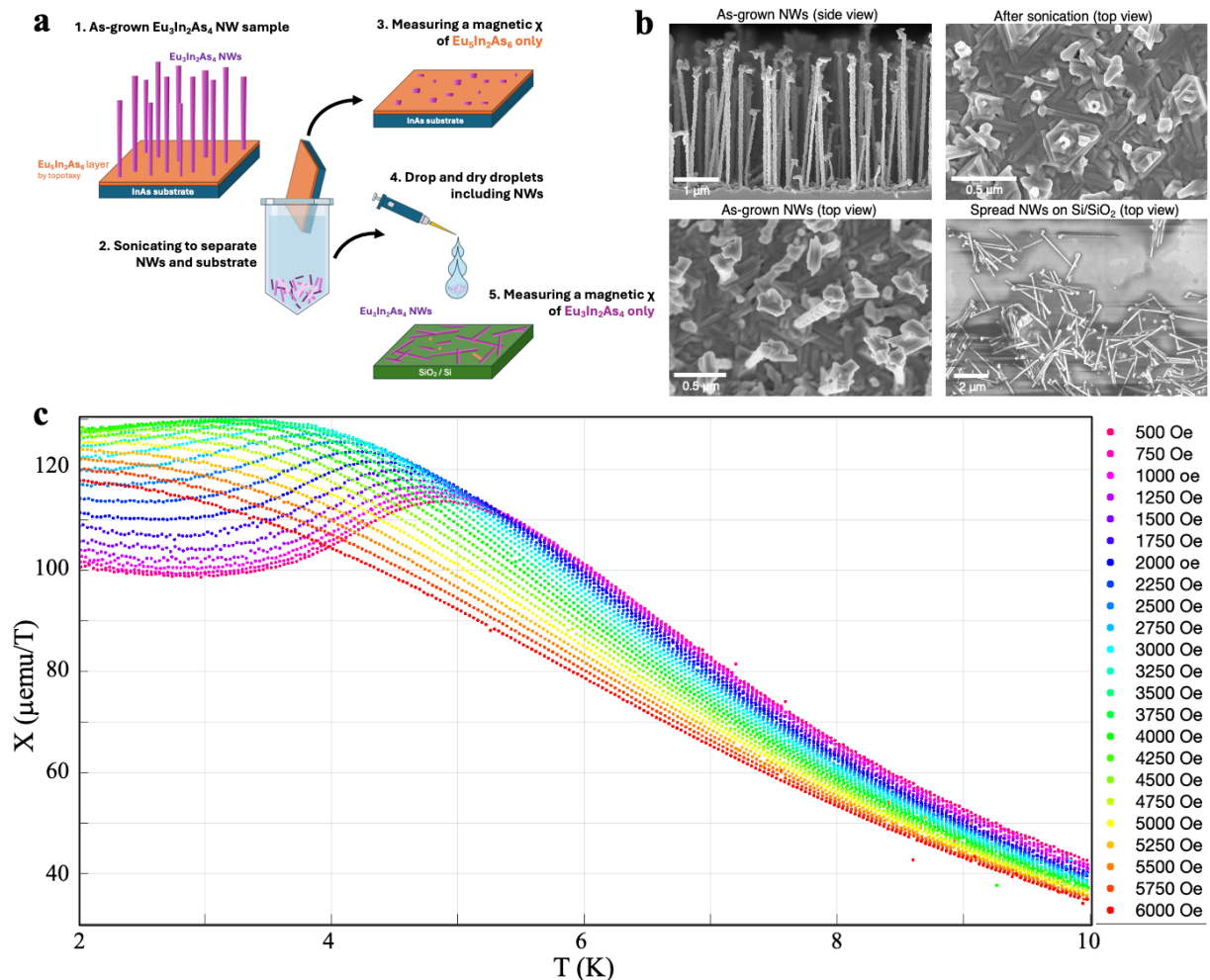

**Supplementary Fig. 14. Global magnetization measurement of $Eu_3In_2As_4$ NWs. (a)** To eliminate any possible magnetic contribution from the InAs substrate, we harvest the NWs by sonicating them for a few seconds in a solvent. We then disperse them over a Si/SiO$_2$ substrate by depositing the suspension and drying it drop by drop to exhaust all the NWs available. **(b)** SEM images of the $Eu_3In_2As_4$ NWs used (top left and bottom left panels), growth substrate after harvesting (top right panel) and the Si/SiO2 substrate after deposition (bottom right). **(c)** Global susceptibility ($\chi$=B/H) measurements of the dispersed $Eu_3In_2As_4$ NWs with a SQUID in a magnetic properties measurements system (MPMS) showing the evolution of the Neel transition with increasing applied magnetic fields yielding the phase diagram in Fig.4d (showing the derivative of the data presented here with respect to temperature).

**Topological Electronic Properties of Eu$_3$In$_2$As$_4$**

Eu$_3$In$_2$As$_4$ has an orthorhombic lattice with the space group of *Pnnm* (No. 58) as shown in Supplementary Fig. 12, with inversion symmetry. The lattice parameters were optimized as $a$=7.0159 Å, $b$=16.6348 Å, and $c$=4.4722 Å, by DFT calculation. These values have +1.6% of a negligible average error, comparable with experimental measurement[4]. To find magnetic ground state, AFM and FM orders parallel to $a$-, $b$-, and $c$-axis were explored, as shown in Supplementary Fig. 13, and magnetic anisotropy energies were calculated (Supplementary Table 1). Only magnetic orders that have a crystallographic primitive cell as a magnetic unit cell are considered. C- and A-type AFM were named by referring to the ordering scheme by Wollan and Koehler[5]. C-type indicates the coupling with antiferromagnetic planes and ferromagnetic order along the perpendicular axis, and A-type denotes the antiferromagnetic stacking of ferromagnetic planes. C-type AFM along $a$-axis (C-AFMa, Supplementary Fig. 13a) was found to be the magnetic ground state of Eu$_3$In$_2$As$_4$. Because antiparallel spin alignment between the nearest Eu atoms lowers the electronic energy, C-type AFM is favored between configurations. AFM hereafter indicates the C-type AFM.

The electronic band structures of Eu$_3$In$_2$As$_4$ in AFMa phase with or without spin-orbit coupling (SOC) are presented in Supplementary Fig. 14. The presence of SOC induces band inversion between In-5$s$ and As-4$p$ is found at Γ (Eg~5 meV). Because the magnetic anisotropy energies in AFM or FM are tiny, about 0.03 meV per Eu atom, we compared the electronic bands around the Fermi energy as shown in Supplementary Fig. 15. AFMa, AFMb, and AFMc phases have an indirect gap of 5 meV, 9 meV, and 21 meV, respectively.

We next examine the topological properties of Eu$_3$In$_2$As$_4$, by computing parity-based symmetry indicator $\mathbb{Z}_4$,

$$\mathbb{Z}_4 = \frac{1}{2}\sum_{a=1}^{8}\sum_{n=1}^{n_{occ}}[1+\xi_n(\Lambda_a)] \mod 4$$

where $\Lambda_a$ are the eight time-reversal invariant momenta (TRIM), n is the band index, $n_{occ}$ is the number of occupied bands, and $\xi_n(\Lambda_a)$ is the parity eigenvalue (±1) of the nth band at $\Lambda_a$. $\mathbb{Z}_4 = 0$ indicates a trivial insulator while $\mathbb{Z}_4 = 1,3$ denotes a Weyl semimetal. $\mathbb{Z}_4 = 2$ corresponds to an axion insulator with a quantized topological magnetoelectric response ($\theta = \pi$). As shown in Supplementary Table 2, the direction of magnetic moments does not change the band order and related topology. Eu$_3$In$_2$As$_4$ is an axion insulator in AFM phase ($\mathbb{Z}_4 = 2$) and a semimetal in FM phase ($\mathbb{Z}_4 = 1$).

For the FMa and FMb phase, Fermi arcs, surface state, and evolutions of the Wannier charge center were calculated as depicted in Supplementary Fig. 17. A pair of Weyl nodes were found on X-Γ-X' ($\mu \parallel \boldsymbol{a}$) and Y-Γ-Y'.

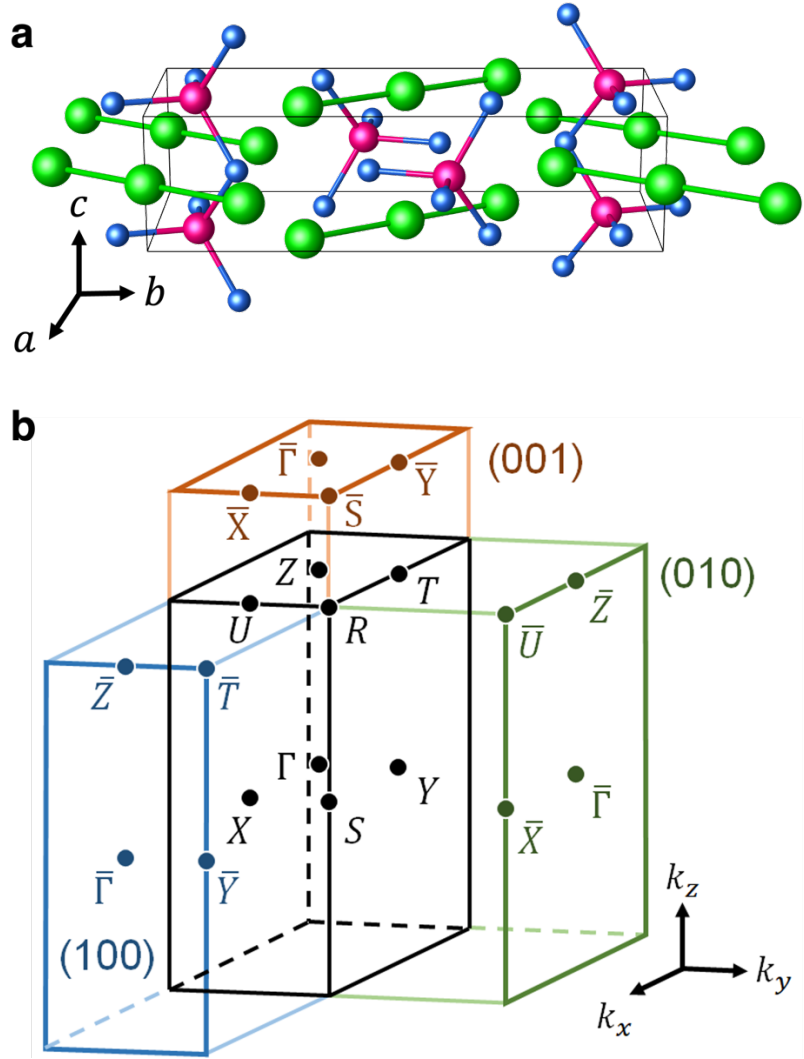

**Supplementary Fig. 15. (a)** Lattice structure of Eu$_3$In$_2$As$_4$. Eu, In, and As atoms are depicted as green, pink, and blue spheres, respectively. Three close Eu atoms with about 4 Å distance are connected to guide eyes. **(b)** The 1$^{st}$ Brillouin zone of orthorhombic cell (black lines) and projected (100), (010), and (001) surface-Brillouin zones with high-symmetry points indicated by capital letters.

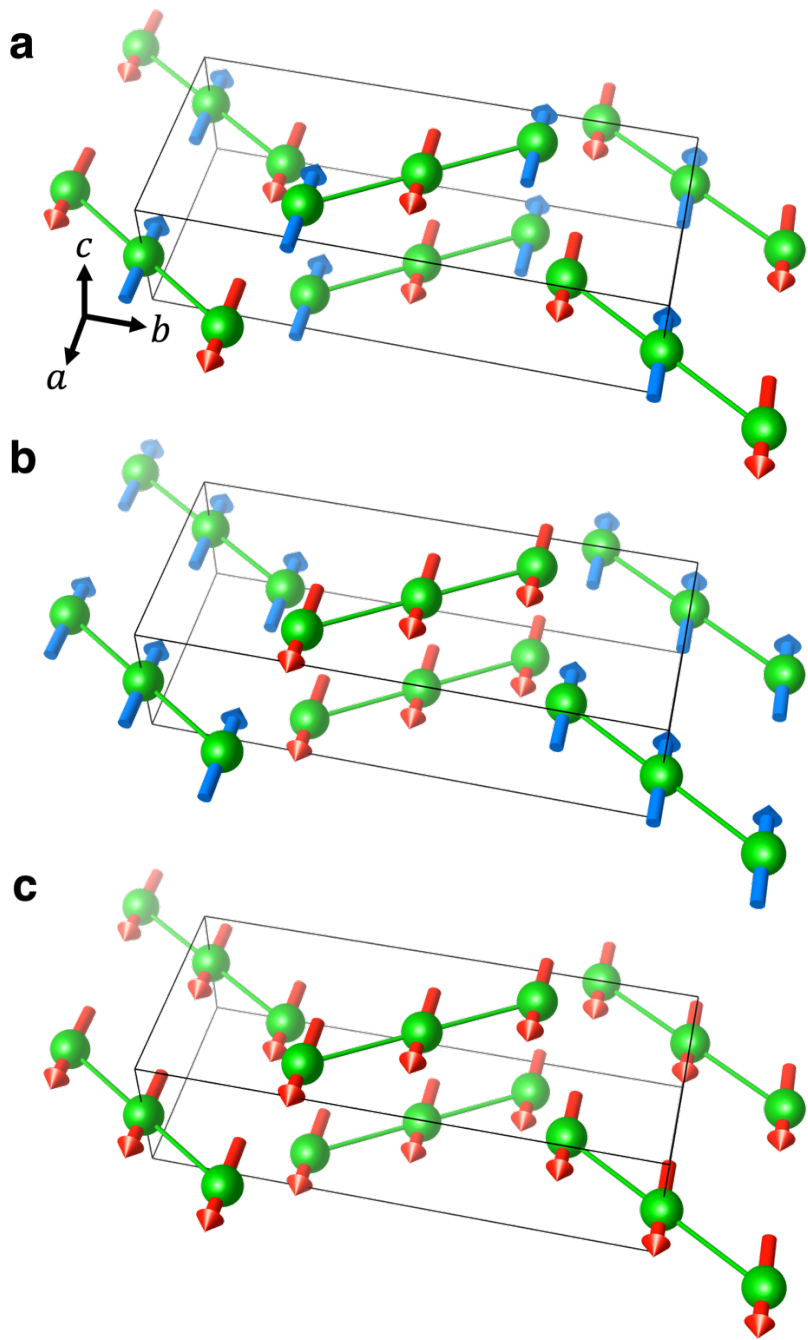

**Supplementary Fig. 16. Magnetic structures in Eu₃In₂As₄ whose magnetic unit cell is its crystallographic primitive cell**. **(a)** C-type and **(b)** A-type antiferromagnetic order, and **(c)** ferromagnetic order according to *a*-axis. Only Eu atoms are illustrated for clarity. Red and blue arrows heading in opposite directions indicate 7 μ$_B$ of spin magnetic moments localized on Eu-4*f* electrons.

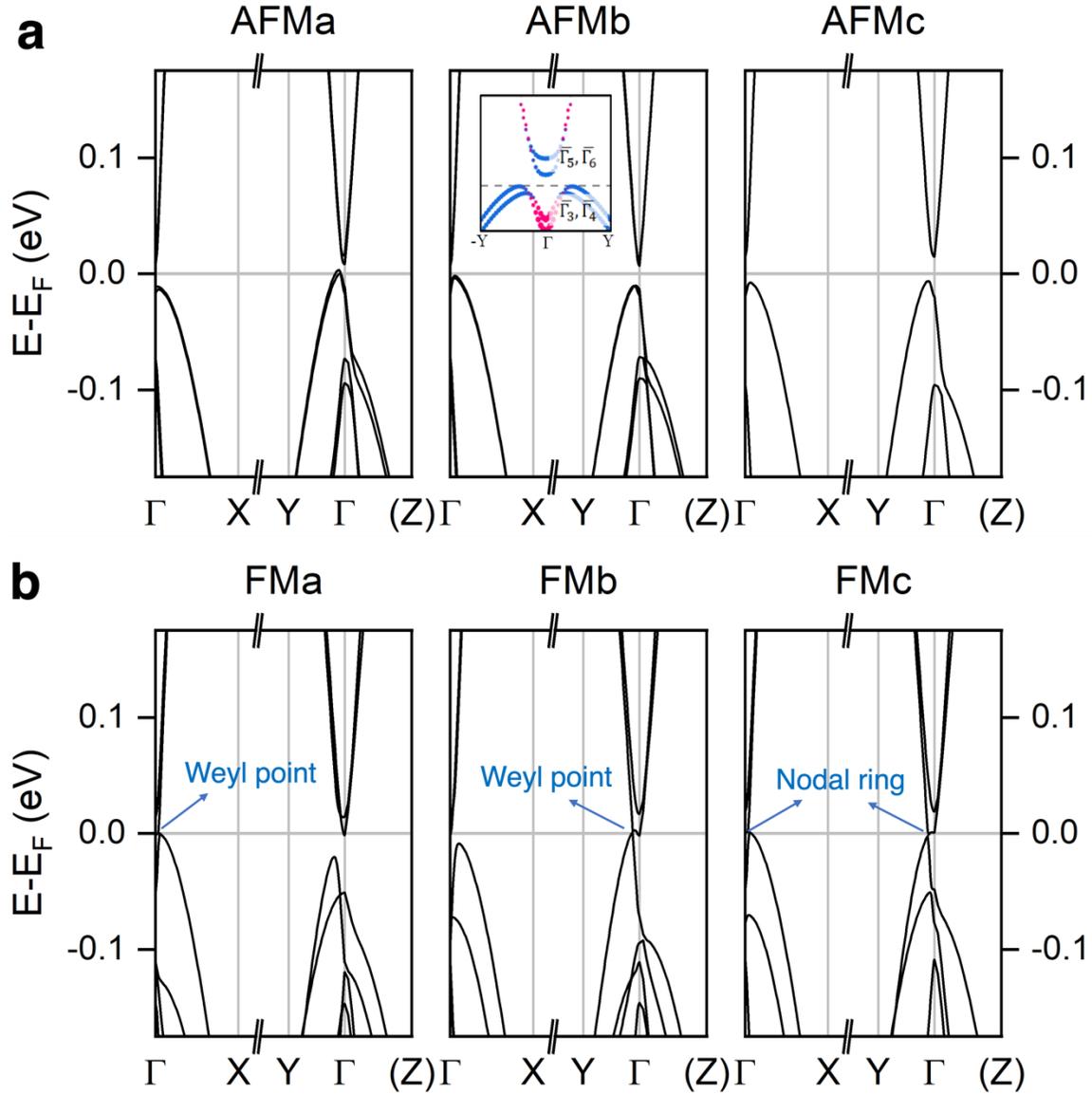

**Supplementary Fig. 17.** Electronic band structures of $Eu_3In_2As_4$ in **(a)** C-type AFM and **(b)** FM phase whose spin magnetic momenta are aligned with *a*-, *b*-, and *c*-axis by considering spin-orbit coupling. All AFM phases exhibit band inversions, presenting axion insulators. The FMa and FMb are Weyl semimetals. But FMc is a Weyl nodal line semimetal where a nodal ring appears around the $\Gamma$ point on the $k_z = 0$ plane. The (Z) means that the direction to Z, and the part of the $\Gamma$-Z path is left out for clarity.

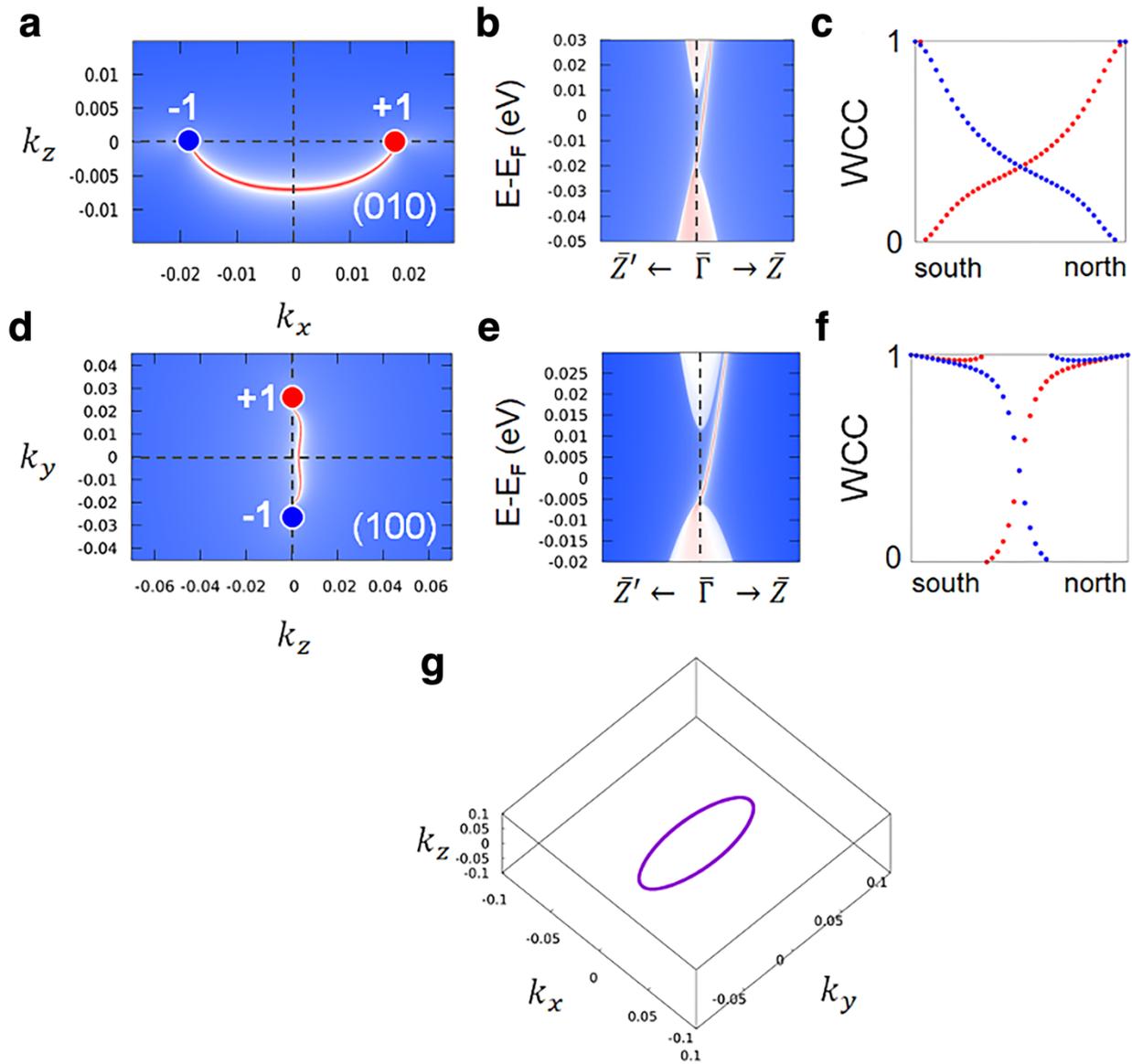

**Supplementary Fig. 18. Fermi arc, surface state, and the motion of the Wannier charge centers (WCC) at Weyl points in the ferromagnetic phase. (a)** Surface states of the (010) termination in FMa phase, confirming the existence of Fermi arc. +1 and -1 with red and blue dots indicate the topological charge of each Weyl points, respectively. **(b)** Surface state spectrum shows a chiral mode. **(c)** Motion of the WCC at two Weyl points. Red and blue dots denote +1/-1 of topological charge which is shown in (a). **(d–f)** Same as (a–c) but in FMb phase. **(g)** Nodal ring at $k_z = 0$ plane of FMc phase.

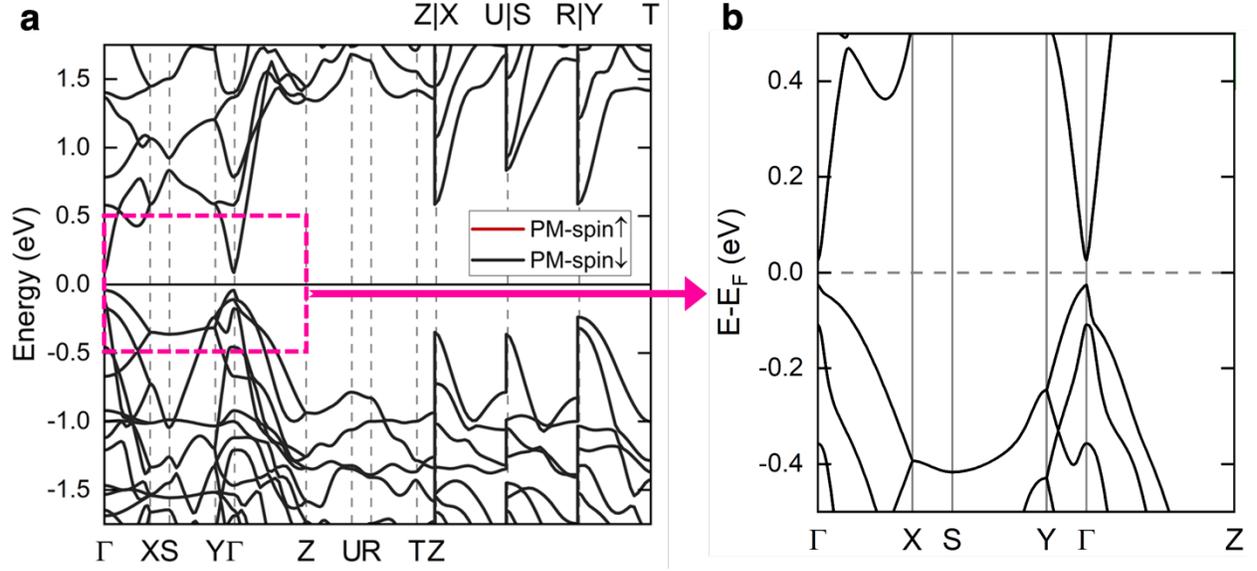

**Supplementary Fig. 19. Electronic band structures of $Eu_3In_2As_4$ in a paramagnetic order.** Each band is double-degenerate guaranteed by the PT symmetry.

**Supplementary Table 1.** Relative energies in different magnetic states of $Eu_3In_2As_4$ (see Supplementary Fig. 13).

| Magnetic order | Energy (meV/Eu atom) |
|---|---|
| AFM (C-type) | 0 (lowest energy) |
| AFM (A-type) | 0.81 |
| FM | 1.69 |

**Supplementary Table 2. Topological invariants for different magnetic phases.** The number of occupied bands of +1/-1 parity eigenvalue at the eight TRIM points, and the total number of electrons is 154. The direction of magnetization does not change the band order and corresponding parity eigenvalues.

| $\Lambda_a$ | $\Gamma(0,0,0)$ | $X(\pi,0,0)$ $Y(0,\pi,0)$ $Z(0,0,\pi)$ | $S(\pi,\pi,0)$ | $U(\pi,0,\pi)$ $T(0,\pi,\pi)$ | $R(\pi,\pi,\pi)$ | Topological index |
|---|---|---|---|---|---|---|
| AFM | 68/86 | 77/77 | 66/88 | 88/66 | 77/77 | $\mathbb{Z}_4 = 2$ for AFMa,b,c |
| FM | 67/87 | 77/77 | 66/88 | 88/66 | 77/77 | $\mathbb{Z}_4 = 1$ for FMa,b,c |